\documentclass[11pt]{article}
\usepackage{epsfig}

\def\thefootnote{\fnsymbol{footnote}}

\newcommand{\lang}{\left\langle}
\newcommand{\rang}{\right\rangle}

\def\bpl{b_+^0}
\def\susy{supersymmetry}
\def\bea{\begin{eqnarray}}
\def\eea{\end{eqnarray}}
\def\beq{\begin{equation}}
\def\eeq{\end{equation}}

\def\w{\bar{w}}
\def\rvev{\right\rangle}
\def\lvev{\left\langle}

\def\dgs{\delta_{GS}}

\def\hka{\hat{\kappa}}

\def\re{{\rm Re}}
\def\im{{\rm Im}}

\def\bM{\bar{M}}
\def\bN{\bar{N}}

\def\W{\overline{W}}

\def\G{{\cal G}}

\def\tmu{{\tilde\mu}}

\def\hK{{\hat K}}

\def\hN{{\hat N}}

\def\pp{\partial}

\def\ibar{\bar{\imath}}

\def\[{\left [}
\def\]{\right ]}
\def\({\left (}
\def\){\right )}
\def\lbr{\left\{}
\def\rbr{\right\}}

\def\T{\bar{T}}
\def\t{\bar{t}}

\def\z{\bar{z}}

\def\S{{\bar{S}}}

\def\Ta{T^\alpha}
\def\bTa{\T^\alpha}
\def\ta{t^\alpha}
\def\bta{\t^\alpha}
\def\na{n^\alpha}
\def\ma{m^\alpha}
\def\hm{\hat{m}}
\def\hma{\hm^\alpha}
\def\pa{p^\alpha}

\def\L{{\cal L}}

\def\hz{\hat{z}}

\def\hh{\hat{h}}

\def\hN{\widehat{N}}

\def\t{\bar{t}}

\def\n{\bar{n}}
\def\m{\bar{m}}

\def\s{\bar{s}}

\def\Z{{\bar{Z}}}

\def\hPh{\hat{\Phi}}
\def\hph{\hat{\phi}}

\def\ts{\tilde{s}}
\def\bF{\bar{F}}

\def\a0{\alpha_0}
\def\const{{1\over32\pi^2}}
\def\mG{m_{3/2}}
\def\up{^{(0)}}
\def\upp{^{(1)}}
\def\U{\bar{U}}
\def\u{\bar{u}}
\def\bts{\bar{\ts}}
\def\vgs{V_{GS}}
\begin{document}

\begin{titlepage} 
\begin{center}
            \hfill    LBNL-46970 \\
            \hfill    UCB-PTH-00/37 \\
            \hfill    LPT-Orsay-00/87 \\
            \hfill hep-ph/0011081\\
        \hfill \today
\vskip .3in
{\large \bf One loop soft supersymmetry breaking terms in 
superstring effective theories}\footnote{This work was supported in part by the
Director, Office of Science, Office of Basic Energy Services,
of the U.S. Department of
Energy under Contract DE-AC03-76SF00098 and in part by the National
Science Foundation under grants PHY-95-14797 and INT-9910077.}\\[.1in]

%Pierre Bin\'etruy, Mary K. Gaillard and Brent D. Nelson

%{\em Department of Physics, University of California, and
%
% Theoretical Physics Group, 50A-5101, Lawrence Berkeley National Laboratory, 
%
%      Berkeley, CA 94720, USA}

Pierre Bin\'etruy, 

{\em LPT, Universit\'e Paris-Sud, Bat.210
F-91405 Orsay Cedex}
\vskip .1in

Mary K. Gaillard and Brent D. Nelson

{\em Department of Physics, University of California, and

 Theoretical Physics Group, 50A-5101, Lawrence Berkeley National Laboratory, 

      Berkeley, CA 94720, USA}

\vskip .2in
\begin{abstract}

We perform a systematic analysis of soft supersymmetry breaking terms
at the one loop level in a large class of string effective field
theories. This includes the so-called anomaly mediated
contributions. We illustrate our results for several classes of
orbifold models. In particular, we discuss a class of models where
soft supersymmetry breaking terms are determined by quasi model
independent anomaly mediated contributions, with possibly
non-vanishing scalar masses at the one loop level. We show that the
latter contribution depends on the detailed prescription of the
regularization process which is assumed to represent the Planck scale
physics of the underlying fundamental theory. The usual anomaly
mediation case with vanishing scalar masses at one loop is not found to
be generic. However gaugino masses and A-terms always vanish at tree
level if supersymmetry breaking is moduli dominated with the moduli
stabilized at self-dual points, whereas the vanishing of the B-term 
depends on the origin of the $\mu$-term in the underlying theory.
We also discuss the supersymmetric spectrum of O-I and
O-II models, as well as a model of gaugino condensation. For
reference, explicit spectra corresponding to a Higgs mass of 114 GeV
are given. Finally, we address general strategies for distinguishing
among these models.

\end{abstract}

\end{center}
\newpage
\pagestyle{empty}
\null
\end{titlepage}
\newpage
\renewcommand{\theequation}{\arabic{section}.\arabic{equation}}
\renewcommand{\thepage}{\arabic{page}}
\setcounter{page}{1}
\def\thefootnote{\arabic{footnote}}
\setcounter{footnote}{0}

\section{Introduction}
In any given supersymmetric theory, a  consistent  analysis 
of the soft terms is necessary in order to make reliable predictions. Such a 
systematic analysis was performed  at tree level by Brignole, Ib\'a\~{n}ez 
and Mu\~{n}oz \cite{bim} some time ago for a large class of four-dimensional 
string models. One of the nice features of this analysis was to make explicit 
the dependence of the soft terms in the auxiliary field vacuum expectation 
values ({\em vev}'s) and thus to relate them directly to the 
supersymmetry breaking mechanism.
In this respect, the auxiliary fields $F_S$ and $F_{T^\alpha}$ associated 
respectively with the string dilaton and the moduli fields are expected to 
play a central role in these superstring models.  

This analysis showed that, besides a universal contribution associated with 
the dilaton field, soft terms generically receive from moduli fields a 
non-universal contribution  which may lead to a very different phenomenology 
from the standard one referred to as the minimal supergravity model.

Recently, a new  contribution to the soft 
supersymmetry breaking terms  has been discussed under the name of 
``anomaly mediated terms'' \cite{rs,hit} that arise at the quantum 
level from the superconformal anomaly. They are truly 
supergravity contributions in the sense that they involve the auxiliary fields 
of the supergravity multiplet, more precisely the complex scalar auxiliary 
field $M$ in the minimal formulation (see {e.g.} \cite{wb} or \cite{physrep}). 
However if these contributions are included, then all one-loop contributions 
to the soft terms should be taken into account. In what follows, we  
present the general form of these contributions, expressed in terms of the 
auxiliary fields  and we discuss them for several classes of superstring 
models. We stress that some of the contributions depend on the way the
underlying theory regulates the low energy effective field theory. In
particular we find a model of anomaly mediation where the scalar
masses might be non-vanishing at one loop.

\section{ General form of one loop  supersymmetry breaking terms}
\setcounter{equation}{0}

In this section, we give the complete expressions for the soft
supersymmetry breaking terms\footnote{We keep only the terms of leading
order in $\mG/\mu_R$, where $\mG$ is the gravitino mass (typically less
than 10 TeV) and $\mu_R$ is the renormalization scale, taken to be the
scale at which supersymmetry is broken (typically $10^{11}$GeV or
higher)}. Let us start by introducing our notations.  We consider a set
of chiral superfields $Z^M$ (the associated scalar field will be denoted
by $z^M$) which belong to two distinct classes: the first class $Z^i$
denotes observable superfields charged under the gauge symmetries, the
second class $Z^n$ describes hidden sector fields, typically in the
models that we will consider the dilaton and T and U moduli fields. Their
interactions are described by three functions: the K\"ahler potential
$K(Z^M, \bar Z^{\bar M})$, the superpotential $W(Z^i, Z^n)$ and the
gauge kinetic functions $f^a(Z^n)$, one for each gauge group $G^a$.

The auxiliary fields are obtained by solving the corresponding equations 
of motion. They read for the chiral superfields:\footnote{ We follow the 
sign conventions of \cite{physrep,bgg}. Let us note that the auxiliary 
fields differ by a sign from the ones used by Brignole, Ib\'a\~nez and Mu\~noz
\cite{bim}.} 
\beq
F^M = - e^{K/2} K^{M \bar N} \left(\bar W_{\bar N} + K_{\bar N} \bar W \right),
\label{F}
\eeq
where, as is standard, $\bar W_{\bar N} = \partial \bar W / \partial 
\bar Z^{\bar N}$ 
and   $K^{M \bar N}$ is the inverse of the K\"ahler metric $K_{M \bar N}=
\partial^2 K / \partial Z^M \partial \bar Z^{\bar N}$. 
The supergravity auxiliary field $M$ simply reads:
\beq
M = -3 e^{K/2} W. \label{M}
\eeq
As a sign of spontaneous breaking of supersymmetry, the gravitino mass is 
directly expressed in terms of its {\em vev} (in reduced Planck scale units 
$M_{Pl}/\sqrt{8\pi} = 1$ which we use from now on):
\beq 
m_{3/2} = -{1 \over 3} < \bar M > = <e^{K/2} \bar W >. \label{m3/2}
\eeq

In terms of these fields,
the $F$-term part of the potential reads:
\beq
V= F^I K_{I \bar J} \bar F^{\bar J} - {1 \over 3} M \bar M. \label{pot}
\eeq
Since in what follows we will assume vanishing $D$-terms we will only be 
interested in this part of the scalar potential. 

Finally, the holomorphic function $f_a(Z^M)$ is the coefficient of the gauge 
kinetic term in superspace. Its {\em vev} yields the gauge coupling associated 
with the gauge group $G_a$: 
\beq 
< \re f_a > = {1 \over g_a^2}. \label{g}
\eeq

In the weak coupling regime, the models that we consider have a simple gauge
kinetic function:
\beq
f^{(0)}_a (Z^n) = k_a S, \label{f=S}
\eeq
where $S$ is the string dilaton and $k_a$ is the affine level\footnote{From 
now on, we will only consider affine level one nonabelian gauge groups 
{\em i.e.} 
$k=1$ ($k=5/3 $ for the abelian group $U(1)_Y$ of the Standard Model).}. In 
what follows, we will adopt the description of the dilaton in terms of a 
chiral superfield, although all our results were obtained in the linear 
superfield formulation as described in Appendix A.
Quantum 
corrections involve the moduli fields $T^\alpha$. Of central importance
at the perturbative level, are the diagonal modular transformations:
\beq
T^\alpha \rightarrow {aT^\alpha - ib \over icT^\alpha +d}, \;\;\; \; ad-bc=1, \;
\;a,b,c,d \in Z, \label{mod}\eeq
that leaves the classical effective supergravity theory invariant.
At the quantum level there is an anomaly~\cite{dkl}--\cite{gs}  which is 
cancelled by a universal Green-Schwarz counterterm \cite{gs} and 
model-dependent string threshold corrections \cite{dkl,ant}. In order to present 
the contributions of these terms to the gaugino masses, we must be 
somewhat more explicit.

We take the standard form:
\beq
K(S,T) = k(S+\bar S) + K(T^\alpha) = k(S+\bar S) - \sum_{\alpha=1}^3 \ln \left(T^\alpha + 
\bar T^\alpha \right), \label{Kahler}
\eeq
for the moduli dependence of the K\"ahler potential. We will assume
for the simplicity of the expressions which follow
that the K\"ahler metric for the matter fields has the form:
\beq
K_{i \bar j} = \kappa_i(Z^n) \delta_{ij} + O(|Z^i|^2). \label{kappa}
\eeq
Indeed a matter field which transforms as
\beq
Z^i \rightarrow   (icT^\alpha + d)^{n_i^\alpha} Z^i
\eeq
under the modular transformations (\ref{mod}) is said to have weight 
$n_i^\alpha$ and has 
\beq \kappa_i = \prod_\alpha (T^\alpha + \bar T^\alpha)^{n^\alpha_i}. 
\label{kap} \eeq
The superpotential transforms as
\begin{equation}
W \rightarrow W \prod_{\alpha} \(icT^{\alpha} + d\)^{-1}.
\end{equation}

\subsection{ Gaugino masses}

The tree level contribution to the  masses of canonically normalized gaugino 
fields simply reads:\footnote{ 
From now on, we will suppress the brackets $<\cdots>$ indicating 
that all  explicit expressions of soft terms are given in terms of {\em 
vevs} of fields.} 
\beq
M^{(0)}_a =  {g_a^2 \over 2}  F^n \partial_n f_a . 
\label{M0}
\eeq 
The full  one loop anomaly-induced contribution has been obtained recently 
\cite{gnw,bmp,gn}. It is:
\beq
M^{(1)}_a|_{\rm an} = {g_a(\mu)^2 \over 2} \left[ {2 b^0_a \over 3} \bar M 
- {1 \over 8 \pi^2} \left( C_a - \sum_i C^i_a \right) F^n  K_ n
- {1 \over 4 \pi^2} \sum_i C^i_a F^n \partial_n  \ln \kappa_i  
\right], \label{Man}
\eeq
where $C_a$, $C_a^i$ are the quadratic Casimir operators for the gauge group 
$G_a$ respectively in the adjoint representation and in the representation 
of $Z^i$, $b^0_a$ is the one loop coefficient of the corresponding 
beta function:
\beq
b^0_a = {1 \over 16 \pi^2} \left( 3 C_a -  \sum_i C_a^i \right),
\label{ba}
\eeq
and the functions $\kappa_i (Z^n)$ have been defined in (\ref{kappa}). The 
first term is the one generally quoted \cite{rs,hit}: $-b^0_a g_a^2  
m_{3/2}$ using  (\ref{m3/2}). 
It is often obtained by a spurion field  
computation \cite{PoRa}.  It is a finite contribution related to the superconformal
anomaly, rather than a remnant of the ultraviolet divergences. 
The remaining terms have been obtained recently \cite{bmp,gn} 
using a general supersymmetric expression for the anomaly-induced terms 
\cite{co} or Pauli-Villars regulators \cite{mkg}.  They reflect the K\"ahler
conformal and chiral anomalies associated with ultraviolet divergences of the
low energy effective field theory~\cite{tom,kl}.

Other terms may appear in string models at one loop. The Green-Schwarz
counterterm has the following form 
\beq
{\cal L}_{\rm GS} = \int d^4\theta E \ L \ V_{{\rm GS}},
\label{GS}\eeq
in a linear multiplet formalism \cite{BDQQ,BGT} where $L$ 
is a linear multiplet which includes the degrees of freedom of the dilaton 
 and of the antisymmetric tensor present among the massless string modes.
The real function $V_{{\rm GS}}$ reads:
\beq V_{{\rm GS}} = {\delta_{GS} \over 24 \pi^2} 
\sum_\alpha \ln \left(T^\alpha + \bar T^\alpha \right) + \sum_i p_i
\prod_\alpha \left(T^\alpha + \bar T^\alpha \right)^{n_i^\alpha} |\phi^i|^2
+ O\left( \phi^4 \right).\label{vgs}\eeq
The group-independent factor $\delta_{GS}$ is simply equal to $-3 C_{E_{_8}}$, 
where $C_{E_{_8}}= 30$ is the Casimir operator of the group $E_8$ in the 
adjoint representation, if there are no Wilson lines. Otherwise, it can be smaller 
in magnitude. 
In the rest of this section, we will neglect\footnote{See Ref. \cite{gnw} for formulas 
taking into 
account the terms of order $\phi^2$.}  terms of order $\phi^2$.

String threshold corrections may be interpreted as one loop corrections to the 
gauge kinetic functions. They read:
\beq
f^{(1)}_a = {1 \over 16 \pi^2} \sum_{\alpha} \ln \eta^2(T^\alpha)
\left[ {\delta_{GS} \over 3} + C_a - \sum_i (1+2\na_i) C^i_a \right],
\eeq
where $\eta(T)$ is the classical Dedekind function:
\beq
\eta(T) = e^{-\pi T /12} \prod_{n=1}^{\infty} (1-e^{-2\pi nT}),
\eeq
which transforms as
\begin{equation}
\eta\(T^{\alpha}\) \rightarrow \(icT^{\alpha} + d\)^{-1/2}
\eta\(T^{\alpha}\)
\end{equation}
under the modular transformation~(\ref{mod}).
We will also use in the following the Riemann zeta function:
\beq
\zeta(T) = {1 \over \eta(T)} {d \eta(T) \over dT}.
\eeq
Combining  contributions from the Green-Schwarz counterterm and string 
threshold corrections with the light loop contribution (\ref{Man})
yields a total one loop contribution \cite{gnw}:
\bea M_a^{(1)} &=& {g_a(\mu)^2 \over 2}\left\{\sum_\alpha F^\alpha
{2 \over 3} \left[{\delta_{GS} \over 16 \pi^2} + b_a^0 - {1 \over 8\pi^2} 
\sum_i C^i_a (1+ 3 \na_i) \right] \left( 2 \zeta(t^\alpha) + 
{1 \over t^\alpha + \bar t^\alpha} \right)\right.\nonumber \\& &\qquad\qquad
\left. + {2 b^0_a \over 3} \bar M + {g_s^2\over16\pi^2}\(C_a - \sum_i
C^i_a\)F^S\rbr.\label{MlGSth}\eea

The last term involves the value of the string coupling at unification. 
In models with  dilaton stabilization through nonperturbative
corrections to the K\"ahler potential \cite{shenk,BaDi}, the value of the gauge 
coupling at the string scale (unification 
scale) $M_s$ is related to $g^2_s = -2K_s$ by:\footnote{ In the linear 
multiplet formulation \cite{BDQQ,BGT}, $g_s^2 = 2 <\ell>$.}
\beq
g^{-2} (M_s) = g_s^{-2} \left( 1 + f(g^2_s/2) \right) =
{\lvev s +\s\rvev\over2}, \label{gstr}
\eeq
where the function $f$ parameterizes nonperturbative string effects \cite{bgw}.

Let us note that the non-holomorphic Eisenstein function 
\begin{equation}
\hat G_2(T, \bar T)
\equiv -2\pi \left(2\zeta(T) + 1/[T+\bar T]\right) \equiv -2\pi G_2(T, \bar T),
\label{Eisenstein}
\end{equation} 
vanishes at the 
self-dual points $T=1$ and $T=e^{i\pi/6}$.

In the presence of the GS term (\ref{GS}), the scalar potential also receives some
corrections.  In particular
\beq K_{M\hN} \to \hK_{M\hN} = K_{M\hN} +
{g_s\over2}\pp_M\pp_{\hN}V_{\rm GS}, \label{khat}
\eeq
in (\ref{F}) and (\ref{pot}). If $p_i=0$, the effect of (\ref{khat})
is to multiply the $vev$ of $F^\alpha$ by the numerical factor
$1 \le 1 - g^2_s\dgs/48\pi^2 \le 1.1$ if $g^2_s = .5$.
Additional corrections are given in Appendix A;
they are unimportant if
$\dgs(\ta + \bta)^{-1}|F^\alpha W_S/W|/48\pi^2 \ll |M/3|$: for example
when  \susy breaking is dilaton dominated or if the superpotential
is independent of the dilaton. The domain of
validity of this approximation is discussed in Appendix A.  We neglect all these corrections 
in the subsequent sections of the text, except in Section 3.5 where $p_i\ne0$ in (\ref{vgs}) is
considered.

%{\bf This paragraph might be better at the end of the introductory part
%of Sect. 2.  But 
%then we would have to move the introduction of the GS term and the
%coupling constant to that section.}

\subsection{A-terms}
A-terms are cubic terms in the scalar potential that generally arise
when supersymmetry is broken:
\beq V_A = {1\over6}\sum_{ijk}A_{ijk}e^{K/2}W_{ijk}z^iz^jz^k + {\rm h.c.} =
  {1\over6}\sum_{ijk}A_{ijk}e^{K/2}(\kappa_i\kappa_j
\kappa_k)^{-{1\over2}}W_{ijk}\hz^i\hz^j\hz^k + {\rm h.c.},\label{aterm} \eeq
where $\hz^i = \kappa_i^{-{1\over2}}z^i$ is a normalized scalar field, and
$W_{ijk} = \pp^3W(z^N)/\pp z^i\pp z^j\pp z^k$.  At tree level we
have
\beq A\up_{ijk} =  \lvev F^n\pp_n\ln(\kappa_i\kappa_j\kappa_k
e^{-K}/W_{ijk})\rvev.\label{aijk}\eeq

The one loop contributions to A-terms (and to scalar masses and B-terms discussed
below) are considerably more sensitive to the details of Planck scale
physics than the gaugino masses considered in the preceding subsection. 
The most straightforward way to regulate an effective theory is by 
introducing heavy fields -- known as Pauli-Villars (PV) fields --
with masses of the order of the effective cut-off, and couplings to
light fields chosen so as to cancel quadratic divergences.  The PV
masses can be interpreted as parameterizing effects of the underlying
theory. These masses are to some extent constrained by supersymmetry.
These constraints are much more powerful in determining the loop-corrected
gaugino masses than the other soft parameters, for the reasons that
follow. 

All gauge-charged PV fields contribute to the vacuum polarization and to
the
gaugino masses.  Their gauge-charge weighted masses are constrained by
finiteness and supersymmetry to give the result in (\ref{MlGSth}).  The
superfield operator that corresponds to these terms is the same one that
contains the field theory chiral and conformal anomalies under K\"ahler
transformations of the type (\ref{mod}), and is therefore completely
determined by the chiral anomaly which is unambiguous. Specifically, the
conformal and chiral anomalies are the real and imaginary part of an
F-term operator; the former is governed by the field dependence of the
PV masses that act as an effective cut-off and are determined by
supersymmetry from the latter~\cite{tom}.

On the other hand, only a subset of charged PV fields $\Phi^A$ 
contribute to the
renormalization of the K\"ahler potential, which determines the matter
wave function renormalization and governs the loop corrections to soft
parameters in the scalar potential.  Their PV masses are determined by
the product of the inverse metrics of these fields 
and of fields $\Pi^A$  to which they couple in the PV superpotential 
to generate Planck scale supersymmetric masses, as well as by 
{\it a priori} unknown holomorphic functions $\mu_A(Z^N)$ of the light
fields that appear in the PV superpotential. 
While the K\"ahler metrics of the $\Phi^A$ are determined by finiteness 
requirements, the metrics of the $\Pi^A$ are arbitrary.  
In operator language, the conformal anomaly
associated with the renormalization of the K\"ahler potential is a
D-term; it is supersymmetric by itself and there is no constraint,
analogous to the conformal/chiral anomaly matching in the case of gauge
field renormalization with an F-term anomaly, on the effective cut-offs
-- or PV masses -- for this term.  As a consequence the
soft terms in the scalar potential cannot be determined
precisely in the absence of a detailed theory of Planck scale physics.

The leading order A-term Lagrangian was given in~\cite{gn}; from
the definition (\ref{aterm}) we obtain for the one loop
contribution:
%\footnote{The explicit expressions given in~\cite{gn} are
%for $\mu_A(Z^N) =$ constant; under this assumption the more general
%expressions given here reduce to those of~\cite{gn}.  Eq. (\ref{aijk1}) is a  
%simplification of the most general expression.  The ``masses'' $m_A$ are
%in fact weighted averages, as displayed explicitly in (32) of~\cite{gn},
%that can differ slightly in different terms in (\ref{aijk1}) and in the
%expression (\ref{smass1}) below for the scalar masses.}
\bea A\upp_{ijk} &=& -{1\over3}\gamma_i\overline{M}
+ \sum_a\gamma_i^a\[2M\up_a\ln(|\hm_im_a|/\mu_R^2) +
F^{n}\pp_{n}\ln(|\hm_im_a|)\]
\nonumber \\ & & 
+ \sum_{lm}\gamma_i^{lm}\[A\up_{ilm}\ln(|m_lm_m|/\mu_R^2) + 
F^{n}\pp_{n}\ln(|m_lm_m|)\] \nonumber\\ & & + \ (i \rightarrow j) \ + \
(j \rightarrow k), \label{aijk1}\eea
where $m_i,m_a$ are the PV masses of the supermultiplets $\Phi^i,\Phi^a$
that regulate loop contributions of the light supermultiplets,
respectively $Z^i,W^\alpha_a$, and $\hm_i$ is the PV mass of a field
$\hPh^i$, in the gauge group representation conjugate to that of $\Phi^i$
(and of $Z^i$) needed to complete the regularization of the gauge-dependent
contribution to the one loop K\"ahler potential
renormalization.\footnote{Assuming a PV mass term  of the form
$\mu_A(Z^N) \Phi^A \Pi^A$ in the superpotential, we have explicitly:
\beq m_A^2 = e^K (\kappa^\Phi_A)^{-1} (\kappa^{\Pi}_A)^{-1}
|\mu_A|^2,\label{PVmass}\eeq 
where $\kappa^\Phi_A$ and  $\kappa^\Pi_A$ are
defined in (\ref{kapPhi})  and (\ref{kapPi}).}
The parameters $\gamma$ determine the chiral multiplet wave function
renormalization.  In the supersymmetric  gauge~\cite{barb} the matter wave
function renormalization matrix is\footnote{ We define the $\gamma$-function
following the conventions of Cheng and  Li~\cite{cl}.} \beq \gamma^j_i =
\const\[4\delta^j_i\sum_ag^2_a(T^2_a)^i_i -  e^K\sum_{kl}W_{ikl}\W^{jkl}\]
.\label{gam}\eeq   The matrix (\ref{gam}) is diagonal
in the approximation in which generation mixing is neglected in the
Yukawa couplings; in practice only the $T^cQ_3H_u$ Yukawa coupling is
important.  We have made this approximation in (\ref{aijk1}), and set 
\bea \gamma_i^j &\approx&\gamma_i\delta^j_i, \quad \gamma_i = 
\sum_{jk}\gamma_i^{jk} + \sum_a\gamma_i^a, \nonumber \\ 
\gamma_i^a &=& {g^2_a\over8\pi^2}(T^2_a)^i_i, \quad \gamma_i^{jk} = -{e^K\over32\pi^2}
(\kappa_i\kappa_j\kappa_k)^{-1}\left|W_{ijk}\right|^2.\label{diag}\eea

We are interested here in string-derived models, in which case the
moduli dependence of the function $W_{ijk}$ is fixed by modular invariance:
\beq
W_{ijk} = w_{ijk} \prod_\alpha \left[ \eta(T^\alpha)
\right]^{2(1+n^\alpha_i+n^\alpha_j+n^\alpha_k)}. \label{WT}
\eeq
Similarly, the quantum
corrected theory should be perturbatively invariant under the modular
transformation (\ref{mod}). This can be achieved if the couplings of the
relevant PV fields are modular invariant.  For the fields
$\Phi^i,\Phi^a,\hPh^i$ that contribute to the renormalization of the K\"ahler
potential, we have~\cite{mkg}, for typical orbifold models,   \beq
\Phi^i:\;\;\kappa^\Phi_i = \kappa_i = \prod_\alpha(T^\alpha + \bTa)^{\na_i},
\quad \hPh^i:\;\; \hka^\Phi_i = \kappa^{-1}_i, \quad \Phi^a: \;\;\kappa^\Phi_a
= g_a^{-2}e^K =  g_a^{-2}e^k\prod_\alpha(T^\alpha +
\bTa)^{-1}.\label{kapPhi}\eeq 
Setting for $\Pi^A =(\Pi^i,  \hat{\Pi}^i, \Pi^a)$, 
\beq \Pi^A:\;\;\kappa^\Pi_A = h_A(S+\S)\prod_\alpha(\Ta +
\bTa)^{\ma_A}, \label{kapPi} \eeq  
the functions $\mu_A(Z^n)$ and therefore
the PV masses  are fixed up to a constant by 
modular covariance, and we obtain for the full A-term, using (\ref{aijk1}),
\bea A_{ijk} &=& {1\over3}A\up_{ijk} - {1\over3}\gamma_i\overline{M} 
- \sum_\alpha F^\alpha\[{1\over\ta + \bta} + 2\zeta(\ta)\]
\(\sum_a\gamma_i^a\pa_{ia} + \sum_{lm}\gamma_i^{lm}\pa_{lm}\)\nonumber \\ & &
+ F^S{\pp\over\pp s}
\(\sum_a\gamma_i^a\ln(\tmu^2_{ia}) + \sum_{lm}\gamma_i^{lm}\ln(\tmu^2_{lm})\)
\nonumber \\ & & 
- \sum_\alpha\ln\[(\ta + \bta)|\eta(\ta)|^4\]\(2
\sum_a\gamma_i^a\pa_{ia}M\up_a + \sum_{lm}\gamma_i^{lm}\pa_{lm}A\up_{ilm}\)
\nonumber \\ & & + 2\sum_a\gamma_i^aM\up_a\ln(\tmu^2_{ia}/\mu_R^2) 
+ \sum_{lm}\gamma_i^{lm}A\up_{ilm}\ln(\tmu^2_{lm}/\mu_R^2) 
 + {\rm cyclic}(ijk), \label{Atot}\eea 
with
\bea \pa_{ij} &=& 1 + {1\over2}\(\na_i + \na_j + \ma_i + \ma_j\), \quad
\pa_{ia} = {1\over2}\(1 + \ma_a + \hma_i - \na_i\), \nonumber \\
\tmu^2_{ij} &=& \mu_i\mu_je^k(h_ih_j)^{-{1\over2}}, \quad \tmu^2_{ai} =
\mu_i\mu_ae^{k/2}g_a(h_a\hh_i)^{-{1\over2}}, \label{params}\eea
where $\mu_i\mu_j$, $\mu_i\mu_a$ are constants.  The tree level A-terms and gaugino masses
are given from (\ref{aijk}) and (\ref{M0}), using (\ref{WT}), respectively by 
\bea A\up_{ijk} &=& \sum_\alpha F^\alpha\(\na_i + \na_j + \na_k + 1\)
\[{1\over\ta + \bta} + 2\zeta(\ta)\] - k_S F^S,
\nonumber \\ M\up_a &=& {g^2_a(\mu)\over2}F^S
= - \pp_s\ln g^2_aF^S.\label{orbma}\eea 

For example, if the PV masses $m_i,m_a,\hm^i$ in (\ref{PVmass}) are constant
(as well as $\mu_A$)\footnote{It was shown in~\cite{mkg} that the K\"ahler
potential for the untwisted sector from orbifold compactification can be made
modular invariant with the relevant masses constant. Since the tree level
K\"ahler potential for the twisted sector is not known beyond quadratic order
in twisted sector fields, the one loop corrections to it cannot be
calculated.}  we have from~(\ref{PVmass})
\bea 
\qquad \qquad \na_i + \ma_i &=& -1, \nonumber \\
(A)\qquad \qquad \na_i - \hma_i &=& 1, \label{A}\\
\qquad \qquad \ma_a &=& 0, \nonumber
\eea
and thus $p^\alpha_{ij} = p^\alpha_{ia} =  0,$ $\tmu^2_{ij}$ and
$\tmu^2_{ai}$ constants.  A commonly (though often implicitly) made assumption in
the literature is instead that $\Pi^A$ has the same K\"ahler metric as
$\Phi^A$: 
\bea
\quad \quad \ma_i &=& \na_i, \nonumber \\
(B) \qquad \qquad \hat m_i^\alpha &=& - \na_i,  \label{B} \\
\quad \quad \ma_a &=& -1; \nonumber
\eea
this gives $\tmu^2_{ij}$ constant, $\tmu^2_{ia}= g^2_a$,  $\pa_{ij} = 1
+ \na_i + \na_j$, $\pa_{ai} = - \na_i$. Distinguishing among
the possibilities from the theoretical point of view requires
string-loop calculations similar to those used to fix the moduli
dependence of the gauge kinetic function~\cite{dkl,ant}.  We note
however that if supersymmetry breaking is moduli mediated ($\langle F^S
\rangle = 0$) with the moduli stabilized at self-dual points, as suggested
by modular invariance, the tree level soft terms (\ref{orbma}) vanish,
and the only one loop contribution is the standard ``anomaly mediated''
term 
\beq A^{\rm anom}_{ijk} = -{1\over3}\overline{M}\(\gamma_i + \gamma_j + \gamma_k\). 
\label{anmed}\eeq  
Therefore if gaugino masses and/or A-terms are measured to be
significantly larger than the ``anomaly mediated'' values (see also
(\ref{MlGSth})], in the string context of assumed modular invariance
this would quite generally
suggest dilaton dominated supersymmetry breaking and/or moduli $vev$'s
far from the self-dual points.

\subsection{B-terms}
B-terms are quadratic terms in $z^i$ and in $\z^{\ibar}$ that appear in 
the scalar potential after supersymmetry breaking if there are such quadratic terms
in the superpotential and/or K\"ahler potential:
\bea W(Z^i) &=& {1\over2}\sum_{ij}\nu_{ij}(Z^n)Z^iZ^j + O[(Z^i)^3], \label{w2}\\
K(Z^i,\Z^{\ibar}) &=& \sum_i\kappa_i|Z^i|^2 + {1\over2}\sum_{ij}\[\alpha_{ij}(Z^n,\Z^{\n})Z^iZ^j 
+ {\rm h.c.}\] + O(|Z^i|^3).\label{k2}\eea
These terms give rise to masses for the chiral supermultiplets $Z^i$:
\bea \L_M &=& - \sum_{ij}\[{1\over2}e^{K/2}\(\psi^i\mu_{ij}\psi^j + {\rm h.c.}\) + e^K
|z^i|^2\kappa^j|\mu_{ij}|^2\], \nonumber \\ \mu_{ij} &=& \nu_{ij} - e^{-K/2}\({1\over3}M
\alpha_{ij} - \bF^{\n}\pp_{\n}\alpha_{ij}\). \label{susymass}\eea
The Lagrangian (\ref{susymass}) is globally supersymmetric although the
mass term 
arising from $\alpha_{ij}$ appears~\cite{giud} only after local
supersymmetry breaking: $m_{3/2}\ne 0$.
The B-term potential takes the form
\beq V_B = {1\over2}\sum_{ij}B_{ij}e^{K/2}\mu_{ij}z^iz^j + {\rm h.c.} =
  {1\over2}\sum_{ij}B_{ij}e^{K/2}(\kappa_i\kappa_j)^{-{1\over2}}\mu_{ij}\hz^i\hz^j 
+ {\rm h.c.}.\label{bterm} \eeq
At tree level we have
\beq B\up_{ij} =  \lvev F^n\pp_n\ln(\kappa_i\kappa_je^{-K}/
\mu_{ij}) + {1\over3}\bM\rvev.\label{bij}\eeq
The one loop contribution is easily extracted from the result for the 
leading order A-term Lagrangian given in~\cite{gn}; we obtain
\bea B\upp_{ij} &=& -{1\over3}\gamma_i\overline{M}
+ \sum_a\gamma_i^a\[2M\up_a\ln(|\hm_im_a|/\mu_R^2) +
F^{n}\pp_{n}\ln(|\hm_im_a|)\] \nonumber \\ & & 
+ \sum_{lm}\gamma_i^{lm}\[A\up_{ilm}\ln(|m_lm_m|/\mu_R^2) + 
F^{n}\pp_{n}\ln(|m_lm_m|)\] + \ (i \rightarrow j).\label{bij1}\eea
Using the assumptions and results of Section 2.2 we obtain for the full 
B-term in string-derived orbifold models
\bea B_{ij} &=& {1\over2}B\up_{ij} - {1\over3}\gamma_i\overline{M} 
- \sum_\alpha F^\alpha\[{1\over\ta + \bta} + 2\zeta(\ta)\]
\(\sum_a\gamma_i^a\pa_{ia} + \sum_{lm}\gamma_i^{lm}\pa_{lm}\)\nonumber \\ & &
+ F^S{\pp\over\pp s}
\(\sum_a\gamma_i^a\ln(\tmu^2_{ia}) + \sum_{lm}\gamma_i^{lm}\ln(\tmu^2_{lm})\)
\nonumber \\ & & 
- \sum_\alpha\ln\[(\ta + \bta)|\eta(\ta)|^4\]\(2
\sum_a\gamma_i^a\pa_{ia}M\up_a + \sum_{lm}\gamma_i^{lm}\pa_{lm}A\up_{ilm}\)
\nonumber \\ & & + 2\sum_a\gamma_i^aM\up_a\ln(\tmu^2_{ia}/\mu_R^2) 
+ \sum_{lm}\gamma_i^{lm}A\up_{ilm}\ln(\tmu^2_{lm}/\mu_R^2) 
 + i\leftrightarrow j, \label{Btot}\eea 
with the various parameters defined in (\ref{params}).  Because we have assumed modular
covariance for trilinear terms in the superpotential,\footnote{In fact we need only 
assume this
for the dominant $T^cQ_3H_u$ term; in making the approximation (\ref{diag}) we implicitly neglect
the small Yukawa couplings that may themselves arise from higher dimension 
operators and/or loop
corrections.} Eqs. (\ref{orbma}) assure that the one loop contribution to the B-term reduces
to the anomaly mediated term 
\beq B^{\rm anom}_{ij} = -{1\over3}\overline{M}\(\gamma_i + \gamma_j\) 
\label{anmedb}\eeq  
 if supersymmetry breaking is moduli mediated ($\langle F^S
\rangle = 0$) with the moduli stabilized at self-dual points.

However tree level B-terms may not vanish in this case; they are sensitive to the origin
of the ``$\mu$-term'' (\ref{susymass}).
A modular invariant K\"ahler potential of the form (\ref{k2}) was constructed~\cite{agnt}
for (2,2) orbifold compactifications of the heterotic string with both T-moduli and 
U-moduli.  Here we restrict the moduli to T-moduli in which case modular
invariance of the K\"ahler potential $K(Z^i,\Z^{\ibar})$ requires
\beq \alpha_{ij}(Z^n,\Z^{\n}) = a_{ij}(S,\S)\prod_\alpha(T^\alpha + \T^\alpha)^{q^\alpha_{ij}}
[\eta(T^\alpha)]^{2k^\alpha_{ij}}[\eta^{*}(\overline{T}^\alpha)]^{2q^\alpha_{ij}},\quad k^\alpha_{ij} = 
q^\alpha_{ij} + \na_i + n_j^\alpha,\label{modk2}\eeq
and modular covariance of the superpotential (\ref{w2}) requires
\beq \nu_{ij}(Z^n) = n_{ij} \prod_\alpha \left[ \eta(T^\alpha)
\right]^{2w^\alpha_{ij}}, \quad w^\alpha_{ij} = 1+n^\alpha_i+n^\alpha_j.\label{nuT}\eeq
Bilinear terms in matter fields do not appear in the tree level superpotential in 
superstring-derived models, but they can be generated from higher
dimension terms when some 
fields acquire $vev$'s.  Bilinear terms in the K\"ahler potential could similarly be
generated from higher dimension terms.  These will be modular invariant if only modular
invariant fields acquire $vev$'s.  For example D-term induced breaking of an anomalous 
$U(1)$ above the scale of supersymmetry breaking preserves modular invariance.  On the other
hand if $\nu_{H_uH_d}\ne 0$, it is of the order of the electroweak
scale: it presumably originates from the {\em vev} $\lvev N \rvev$ of an
electroweak singlet field $N$ and there is no reason that modular
invariance should still be operative at such low energy scales. In any
case, the corresponding B-term is generated by an A-term in this instance.

%phenomenological constraints requires that it be less than about
%a $TeV$, suggesting that it is induced by the $vev$ of a Standard Model singlet $N$,
%$\lvev n = N|\rvev\sim \mu_{H_uH_d}\sim TeV$ at a scale well below 
%the supersymmetry breaking
%scale, typically larger that $10^{12}$ TeV, at which the moduli $vev$'s are presumably
%fixed.  In this case one would not expect ***????? **** modular invariance to be respected.
%On the other hand, to correctly treat such a scenario one should first consider the A-term
%generated at the \susy-breaking scale by the $NH_uH_d$ term in the superpotential, which is
%determined as in Section 2.2.  A B-term then develops at a much lower scale when $N$
%acquires a $vev$;  in this case the tree level B-term is similar to the A-terms, and
%vanishes under the same conditions.\footnote{One cannot get the same result by setting
%$N\to\lvev N\rvev$ in (\ref{bij}) because that equation was obtained under the assumption
%$\lvev Z^i \rvev=0.$}

To consider the case in which the $\mu$-term is already present at the
supersymmetry breaking scale, we can parameterize $\alpha_{ij},\nu_{ij}$ as in
(\ref{modk2}) and (\ref{nuT}), but with the exponents
$k^\alpha_{ij},q^\alpha_{ij},
w^\alpha_{ij}$ left {\it a priori} arbitrary; the case
of modular invariance is recovered when the last equalities in those
equations are imposed.  We also assume that Standard Model singlets $N$ whose
$vev$'s may generate quadratic terms in the superpotential or K\"ahler potential
do not contribute to \susy-breaking: $F^N=0$.

If the $\mu$-term (\ref{susymass}) is generated
by a superpotential term (\ref{w2}), we obtain for the tree level B-term
\bea \[B\up_{ij}\]_{\rm superpotential} &=& \sum_\alpha F^\alpha\[\(1 + \na_i + \na_j 
\){1\over t^\alpha +\t^\alpha}
+ 2\zeta(\ta)w^\alpha_{ij}\] - k_SF^S + {1\over3}\bM.\label{orbbw}\eea
The coefficients of the moduli auxiliary fields vanish at 
the moduli self-dual points when modular invariance (\ref{nuT}) is
imposed, but the B-term does not vanish: $B\up= {1\over3}\bM$ for $F^S = 0$.  
Although it seems rather implausible that a
hierarchically small value of $\nu_{ij}\le$ TeV would be generated at the
supersymmetry breaking scale $\ge 10^{11}$, it could conceivably arise as a
product of $vev$'s
in a superpotential term of very high dimension~\cite{joel}.  

A more natural
origin for a $\mu$-term of the order of a TeV is a quadratic term in the K\"ahler potential as in
(\ref{k2}). The expression for $B\up$ obtained from the general parameterization (\ref{modk2})
is rather complicated and does not in general vanish when $\lvev F^S\rvev = 0$
and modular invariance~(\ref{modk2}) is imposed.  As an example, consider the
simplifying assumptions that $a_{ij}(S,\S) =$ constant and $q^\alpha_{ij}= 
\lvev\pp W/\pp\ta\rvev =0$, then for $\nu_{ij}=0$
\bea \mu_{ij} &=& a_{ij}W[\eta(\ta)]^{2k^\alpha_{ij}}, \quad F^\alpha = 
-{1\over3}\(\ta + \bta\), \nonumber \\
\[B\up_{ij}\]_{\rm K\ddot{a}hler\; potential} &=& \sum_\alpha F^\alpha\[\(1 + \na_i + \na_j 
\){1\over t^\alpha +\t^\alpha} + 2\zeta(\ta)k^\alpha_{ij}\] \nonumber \\ & &
- \(k_S + \pp_S\ln W\)F^S + {1\over3}\bM. \label{orbbk}\eea 
In this case even the coefficients of the moduli auxiliary fields do not vanish at 
the moduli self-dual points when modular invariance is imposed, and 
under the above conditions we get $B\up_{ij} = -{2\over3}\bM$.
It is possible that a comparison of $B_{H_uH_d}$ with A-terms 
might shed some light on the origin of the $\mu$-term (\ref{susymass}).

\subsection{Scalar masses}

The expression ``soft scalar masses'' refers to mass terms in the
scalar potential
\beq V_M = \sum_iM^2_i\kappa_i|z^i|^2= \sum_iM^2_i|\hz^i|^2, \eeq
with no supersymmetric counterpart in the chiral fermion Lagrangian.
The tree level soft scalar masses are given by
\beq (M\up_i)^2 = {1\over9}M\bM - F^n\bF^{\m}\pp_n\pp_{\m}\ln\kappa_i.
\label{smass}\eeq
Here and throughout the discussion of scalar masses, we drop terms
proportional to the vacuum energy, Eq. (\ref{pot}). 

The one loop contribution~\cite{gn} to soft masses is determined by the
soft parameters of the  PV sector.  The A-terms of the PV sector
and the masses of $\phi^i,\phi^a,\hph^i$ are determined by the soft
parameters of the light field tree Lagrangian.  Denoting by $N_A$ the
soft mass of $\pi^A,$ the one loop scalar masses
can be written in the form
\bea (M\upp_i)^2 &=& - {1\over2}\[\sum_a\gamma^a_i\(N_a^2 + \hN_i^2 -
(M\up_a)^2 - (M\up_i)^2\) + \sum_{jk}\gamma^{jk}_i\(N_j^2 + N_k^2 + 
(M\up_j)^2 + (M\up_k)^2\)\] \nonumber \\ & &
- \sum_a\gamma^a_i\[3(M\up_a)^2 - (M\up_i)^2 + M\up_a\(\bF^{\m}\pp_{\m} + 
F^n\pp_n\)\]\ln(|\hm_im_a|/\mu_R^2) \nonumber \\ & &
- \sum_{jk}\gamma^{jk}_i\[(M\up_j)^2 + (M\up_k)^2 + (A\up_{ijk})^2
+ {1\over2}A\up_{ijk}\(\bF^{\m}\pp_{\m} + F^n\pp_n\)\]\ln(|m_jm_k|/\mu_R^2) 
\nonumber \\ & & + {1\over3}\(M + \bM\)\[\sum_a\gamma^a_i M\up_a +
{1\over2}\sum_{jk}\gamma^{jk}_iA\up_{ijk}\]. \label{smass1}\eea

For orbifold compactifications of string theory, with the K\"ahler
metrics given in~(\ref{kap}), we obtain for the tree level scalar masses
\beq (M\up_i)^2 = {1\over9}M\bM + \sum_\alpha F^\alpha\bF^\alpha\na_i
(\ta + \bta)^{-2}.\label{orbsm}\eeq
Note that if $\langle \pp W/\pp t\rangle = 0$, 
then $F^\alpha =
-{1\over3}\bM(\ta + \bta)$ and $M\up_i = 0$ in the no-scale
case with $\sum_\alpha\na_i = -1$, as for the untwisted sector of
orbifold models.  The soft masses $N_A$ are given by the standard
formula (\ref{smass}) by just replacing $\kappa_i$ by $\kappa_A$.
The one loop contribution is then given by
\bea (M\upp_i)^2 &=& {1\over9}M\bM\gamma_i - F^S\bF^{\S}\pp_s\pp_{\s}
\(\sum_a\gamma^a_i\ln \tmu^2_{ia} +  \sum_{jk}\gamma^{jk}_i\ln \tmu^2_{jk}\)
\nonumber \\ & & - \sum_\alpha F^\alpha\bF^\alpha(\ta + \bta)^{-2}
\(\sum_a\gamma^a_ip^\alpha_{ai} + \sum_{jk}\gamma^{jk}_ip^\alpha_{jk}\)
\nonumber \\ & & + {1\over3}\(M + \bM\)\[\sum_a\gamma^a_i M\up_a +
{1\over2}\sum_{jk}\gamma^{jk}_iA\up_{ijk}\] \nonumber \\ & & 
+ \lbr\sum_\alpha F^\alpha\[{1\over\ta + \bta} + 2\zeta(\ta)\]
\(\sum_a\gamma_i^a\pa_{ia}M\up_a + {1\over2}\sum_{jk}\gamma_i^{jk}\pa_{jk}
A\up_{jk}\) + {\rm h.c.}\rbr\nonumber \\ & &
+ \lbr F^S{\pp\over\pp s}\(\sum_a\gamma_i^aM\up_a\ln(\tmu^2_{ia}) + 
{1\over2}\sum_{jk}\gamma_i^{jk}A_{ijk}\up\ln(\tmu^2_{jk})\) + {\rm h.c.}\rbr
\nonumber \\ & & 
- \sum_\alpha\ln\[(\ta + \bta)|\eta(\ta)|^4\]\lbr\sum_a\gamma_i^a\pa_{ia}
\[3(M\up_a)^2 - (M\up_i)^2\]\right. \nonumber \\ & & 
\nonumber \\ & &  \qquad\qquad +
\left.\sum_{jk}\gamma_i^{jk}\pa_{jk}\[(M\up_j)^2 
+ (M\up_k)^2 + (A\up_{ijk})^2\]\rbr
\nonumber \\ & & + \sum_a\gamma_i^a\[3(M\up_a)^2 - (M\up_i)^2\]
\ln(\tmu^2_{ia}/\mu_R^2) \nonumber \\ & & + \sum_{jk}\gamma_i^{jk}\[(M\up_j)^2 
+ (M\up_k)^2 + (A\up_{ijk})^2\]\ln(\tmu^2_{jk}/\mu_R^2) \label{orbsm1}\eea 

%*** This is clearly a mess.  Maybe it simplifies in specific models that
%predict the tree-level terms?  Also maybe we can neglect nonpert. string
%effects in the loop corrections, in which case it would be rational to
%assume $\tmu^2_{AB} = \mu_A\mu_B(s + \s)^{q_{AB}}$, in which case the $F^S$ terms
%should be linear in the $q$'s just as the $F^T$ terms are linear in the
%$p$'s.  Then maybe under certain assumptions (moduli dominated or
%dilaton dominated or certain models?) we could devise some sum rules
%among A-terms and soft masses?***

\section{Orbifold models}
\setcounter{equation}{0}

Following \cite{bim}, we will consider models where the supersymmetry breaking 
arises through non-vanishing expectation values of the auxiliary fields 
$F^S$, $F^\alpha$ and $M$ and we write:
\begin{eqnarray}
 F^S &=& {1 \over \sqrt{3}}\bar M K_{S\bar S}^{-1/2} \sin \theta e^{-i
\gamma_S}, \label{sin} \\
F^\alpha &=& {1 \over \sqrt{3}}\bar M K_{\alpha \bar \alpha}^{-1/2} \cos
\theta \ \Theta_\alpha e^{-i \gamma_\alpha}, \label{alphacos}
\end{eqnarray}
with $\sum_\alpha \Theta_\alpha^2 = 1$. In the case where one considers a 
single common modulus $T$ (the overall radius of compactification), 
(\ref{alphacos}) simply reads:
\begin{equation}
F^T = {1 \over \sqrt{3}}\bar M K_{T \bar T}^{-1/2} \cos \theta 
 e^{-i \gamma_T}. \label{cos}
\end{equation}
Note that the {\em vev} of $M$ 
is related to the gravitino mass through (\ref{m3/2}) and that these auxiliary 
fields automatically satisfy the constraint that the potential $V$ in 
(\ref{pot}) vanishes at the ground state.

%In the following survey of the low-energy phenomenology of these
%orbifold models we will neglect complications which arise from kinetic
%mixing between the moduli induced by the presence of a Green-Schwarz
%counterterm. 
In contrast with Ref.~\cite{bim},  we have already
included the effect of the Green-Schwarz term on the scalar potential
at tree level
and thus the auxiliary fields considered here include to a large extent 
the corresponding Green-Schwarz corrections. Additional corrections (see
(\ref{khat}) and following text) will be discussed in Appendix~A.

In the orbifold models that we consider, {\em i.e.} with gauge kinetic
function~(\ref{f=S}) and K\"ahler potential given by~(\ref{Kahler})
and~(\ref{kap}), the tree level soft terms have simple expressions:
\begin{eqnarray}
M_{a}^{(0)}&=&\frac{g_{a}^{2}}{2\sqrt{3}}\overline{M}k_{s\overline{s}}^{-1/2}
\sin\theta e^{-i \gamma_{S}} \nonumber \\
A^{(0)}_{ijk}&=&\frac{\overline{M}}{\sqrt{3}} \lbr \cos\theta  \sum_{\alpha} 
  (t^{\alpha}+
\overline{t}^{\alpha}) G^{\alpha}_{2} \Theta_{\alpha} (n^{\alpha}_{i}+
n^{\alpha}_{j} + n^{\alpha}_{k} +1) e^{-i \gamma_{\alpha}}
-\frac{k_{s}}{k_{s\overline{s}}^{1/2}} \sin\theta e^{-i \gamma_{S}}
\rbr \nonumber \\
B_{ij}^{(0)}&=& \frac{\overline{M}}{\sqrt{3}}\lbr \frac{1}{\sqrt{3}}
-\frac{\sin\theta}{k_{s\overline{s}}^{1/2}} \[k_{s} + \partial_{s}\ln
\mu_{ij}\] e^{-i\gamma_{S}} +\cos\theta \sum_{\alpha}
\Theta_{\alpha} \[(n_{i}^{\alpha} + n_{j}^{\alpha} +1)
-\partial_{t_{\alpha}} \ln \mu_{ij} \] e^{-i\gamma_{\alpha}} \rbr
\nonumber \\
(M_{i}^{(0)})^{2}&=&\frac{M\overline{M}}{9} \lbr 1+3\sum_{\alpha}n^{\alpha}_{i}
\Theta_{\alpha}^{2}\cos^{2}\theta \rbr
\label{treeterms}
\end{eqnarray}
The one loop contributions to~(\ref{treeterms}) are decidedly more
cumbersome and the complete expressions are given in
Appendix~B. Below
we consider the phenomenological implications of some specific cases
in which the soft supersymmetry breaking terms are simpler. In all of
the following $G^{\alpha}_{2}=(2\zeta(T^{\alpha})
+1/(T^{\alpha} +\overline{T}^{\alpha}))$, which is proportional to the
Eisenstein function~(\ref{Eisenstein}).

\subsection{Moduli domination at the self-dual point: the case for
leading anomaly-induced contributions}
\label{sec:anomalycase}

The analysis of the preceding sections indicates a very specific
situation which turns out to give quasi-model independent contributions.
It is the case of moduli mediated supersymmetry breaking ($F^S=0$ or $\theta=0$) where the
moduli fields lie at a self dual point ($t^\alpha = 1$ or $e^{i\pi/6}$, and
thus  $ G^{\alpha}_{2} =0 $). Assuming
(\ref{f=S}), we have vanishing tree level gaugino masses and A-terms and
from (\ref{MlGSth}), (\ref{Atot}) and~(\ref{anmedb}) we obtain:
\begin{eqnarray}
M_a  &=& g_a(\mu)^2 \ {b^0_a \over 3} \bar M, \\
A_{ijk} &=& -{1 \over 3} \overline{M} (\gamma_i + \gamma_j +
\gamma_k). \label{anomAterm} \\
B_{ij} &=& -\frac{1}{3} \overline{M} (\gamma_i + \gamma_j)
\end{eqnarray}
Further assuming that  $\Theta^2_\alpha ={1\over3}$ (as in the case of
a single modulus $T$, see (\ref{cos})), $\gamma_\alpha=0$
and  $\sum_\alpha\na_i= -1$ (as in the untwisted sector), we have vanishing 
tree level scalar masses and
\beq
M^2_i = {1 \over 9} M \bar M \left[ \gamma_i - \sum_{\alpha,a} \gamma^a_i
p^\alpha_{ai} - \sum_{\alpha,jk} \gamma^{jk}_i p^\alpha_{jk} \right].
\label{anomMass}
\eeq
For the choice (A) of PV weights (see (\ref{A})), one finds 
$M^2_i = M \bar M \gamma_i /9$ 
whereas for the choice (B)  (see (\ref{B})), one obtains
$M^2_i = 0$. This shows very clearly how dependent the
scalar masses are on the regularization scheme forced upon us by the underlying
theory. Case (B) corresponds to what is usually referred to as the
anomaly mediated scenario in which scalar mass-squareds arise at two loops
but are negative for sleptons, thus implying an unacceptable
phenomenology without further {\it ad hoc} assumptions. As
discussed in Section 2.4, if the $\mu$-term
(\ref{susymass}) has a low energy origin through the $vev$ of a standard model
singlet in a superpotential term, we would expect that in this scenario the B-term would
also be dominated by the anomaly mediated contribution (\ref{anmedb}).  On the other hand
if it arises from Planck-scale physics, we do not expect the tree level contribution to
vanish.

Let us note moreover that any departure from our hypothesis ({\em i.e.} a small
value for $F^S$ or a departure from the self-dual point in moduli space)
generates tree level values for the soft terms which tend to overcome the
one loop anomaly-induced contributions considered here, as we will see in the 
next subsection.

When~(\ref{anomMass}) represents the leading contribution to scalar
masses we can see from~(\ref{diag}) that the positivity of scalar
mass squared depends on the size of the Yukawa couplings (which themselves
are a function of the value of $\tan\beta$ and of the scale
$\Lambda_{\rm UV}$ at which the
soft terms are determined) and the values of
the high-scale parameters $p_{ia}^{\alpha}$ and $p_{ij}^{\alpha}$
of~(\ref{params}). In the simple case of scenario (A)~(\ref{A})
mentioned above,
the sign of the scalar mass-squared depends on the sign of the
anomalous dimensions. Keeping all third generation Yukawa couplings
and taking the running masses of the third generation fermions at the Z-mass
to be $\lbr m_{t}, m_{b}, m_{\tau} \rbr = \lbr165, 4.1, 1.78 \rbr$ GeV, we
investigated the range in $\tan\beta$ for which the scalar masses are
positive for a GUT-inspired boundary scale of $\Lambda_{\rm UV} = 2
\times 10^{16}$ GeV
as well as an intermediate scale of $\Lambda_{\rm UV} = 1 \times
10^{11}$ GeV. As can be
seen from Table~\ref{tbl:tanBbounds} the problem of tachyonic scalar
masses for the matter fields is eased considerably in this scenario
relative to the
previously studied anomaly mediated scenario represented by case~(B)~(\ref{B}).

\begin{table}[th]
{\begin{center}
\begin{tabular}{|ccc|} \cline{1-3}
Scalar Mass & $\Lambda_{\rm UV} = 1 \times 10^{11}$ GeV &
$\Lambda_{\rm UV} = 2 \times 10^{16}$ GeV \\ \cline{1-3}
$M_{Q_3}^{2}$ & $1.4 \leq \tan\beta \leq 45$ & $1.7 \leq \tan\beta \leq 44$ \\
$M_{U_3}^{2}$ & $1.8 \leq \tan\beta \leq 48$ & $1.9 \leq \tan\beta \leq 44$ \\
$M_{D_3}^{2}$ & $1.3 \leq \tan\beta \leq 42$ & $1.6 \leq \tan\beta \leq 41$ \\
$M_{L_3}^{2}$ & $1.3 \leq \tan\beta \leq 46$ & $1.6 \leq \tan\beta \leq 44$ \\
$M_{E_3}^{2}$ & $1.3 \leq \tan\beta \leq 39$ & $1.6 \leq \tan\beta \leq 41$ \\
$M_{H_{u}}^{2}$ & always negative & $3.6 \leq \tan\beta \leq 33$ \\
$M_{H_{d}}^{2}$ & $1.3 \leq \tan\beta \leq 33$ & $1.6 \leq \tan\beta \leq 37$ \\
\cline{1-3}
\end{tabular}
\end{center}
{\caption{{\footnotesize {\bf Regions of Positive Mass-Squared in the
        Anomaly Dominated Scenario.} Range of $\tan\beta$ for which
      scalar mass-squareds are positive at the boundary scale
      $\Lambda_{\rm UV}$ using the PV scenario~(A). The value of
  $\tan\beta$ was varied over the range for which the third generation
  Yukawa couplings remain perturbative up to the scale $\Lambda_{\rm UV}$. This
  corresponds to the range $1.3 \leq \tan\beta \leq 44$ for
  $\Lambda_{\rm UV} = 2 \times 10^{16}$ GeV and $1.6 \leq \tan\beta \leq 48$ for
  $\Lambda_{\rm UV} = 1 \times 10^{11}$ GeV.} }}}
\label{tbl:tanBbounds}
\end{table}

\begin{figure}[t]
    \begin{center}
\centerline{
       \epsfig{file=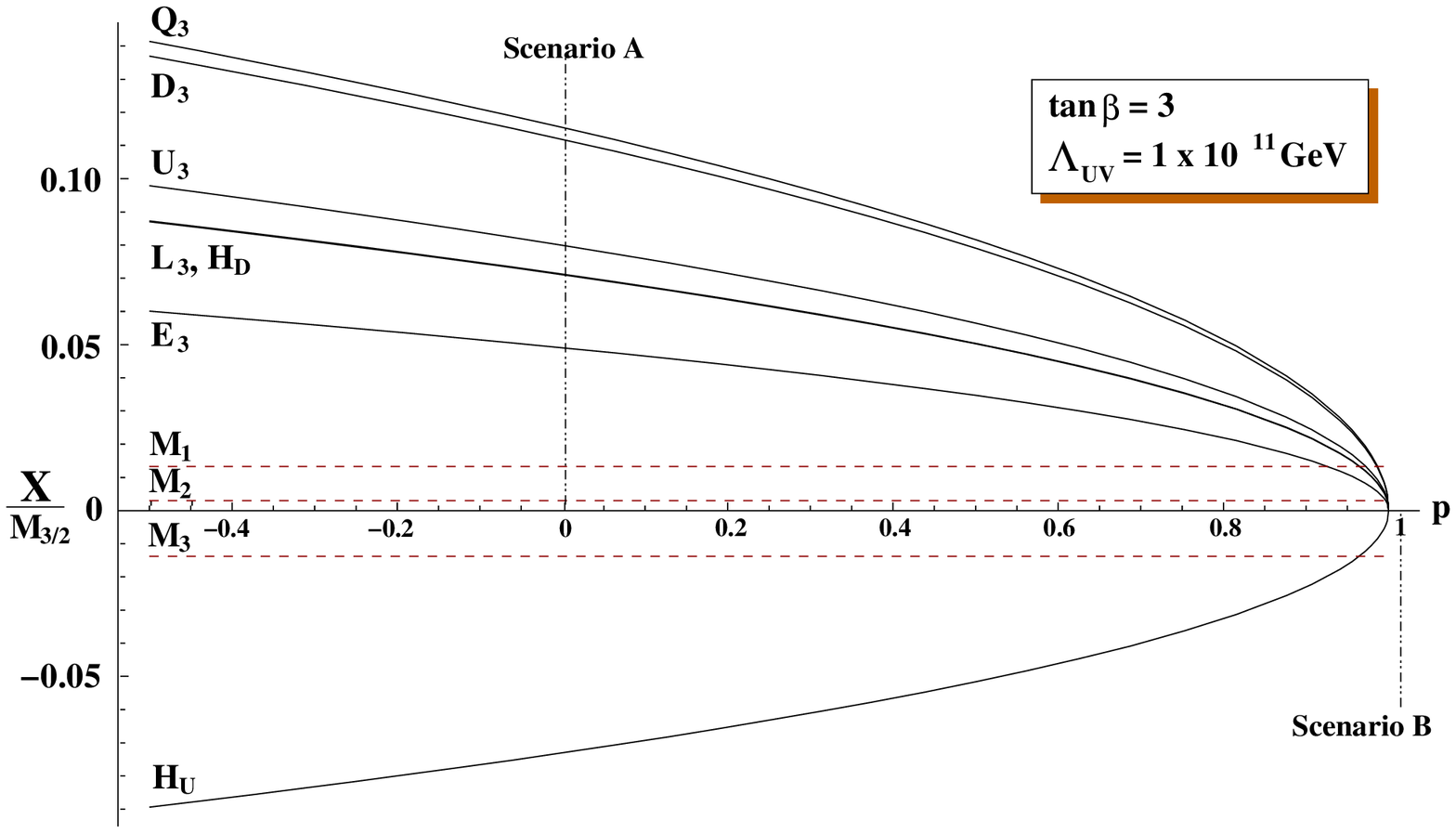,width=0.5\textwidth}
       \epsfig{file=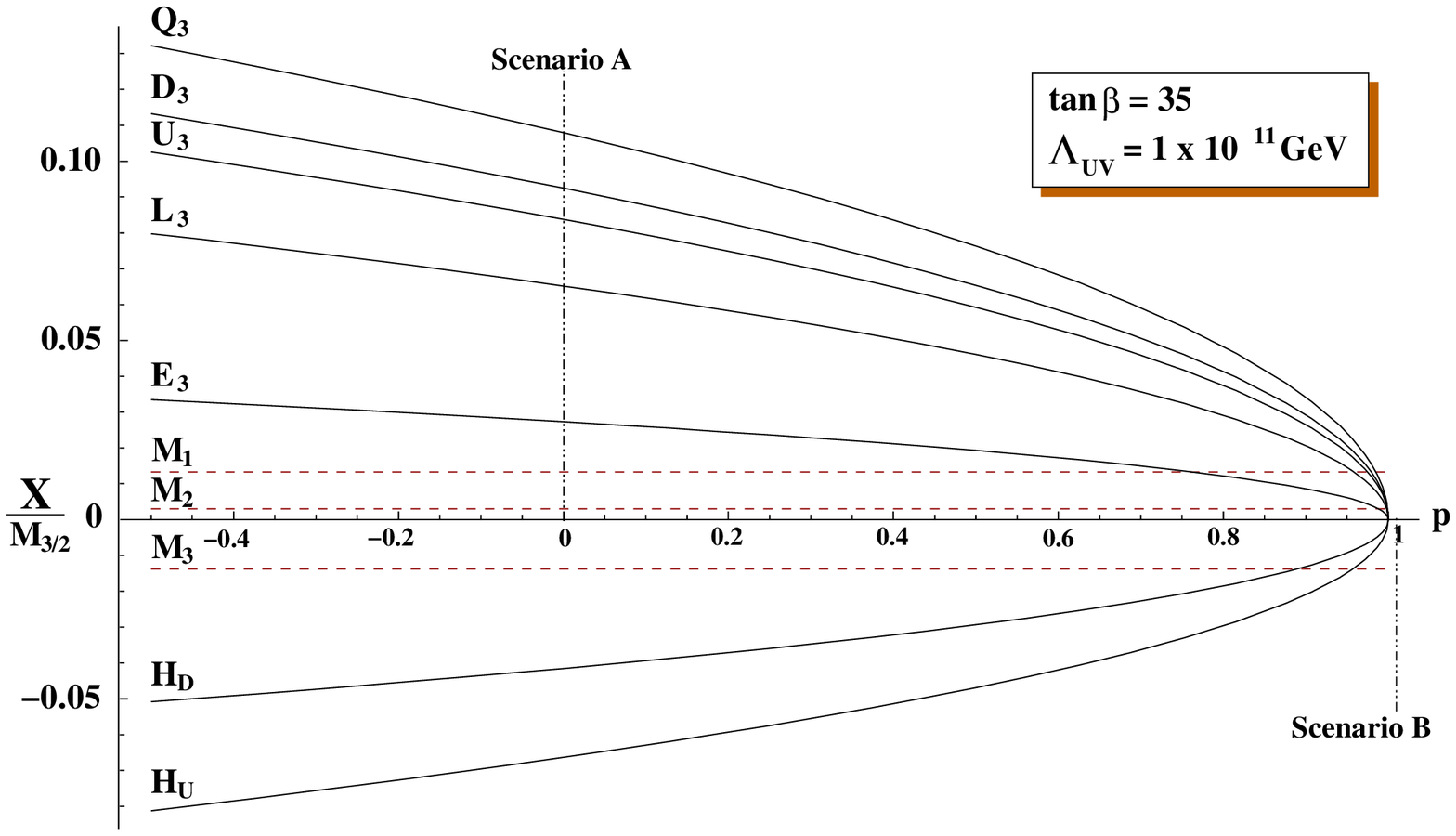,width=0.5\textwidth}}
          \caption{{\footnotesize {\bf Soft Term Spectrum for
                Anomaly Dominated Scenario}. Soft term magnitudes
              for third generation scalars, Higgs fields and gaugino
              masses are given as a function of universal PV parameter
              $p$ as a fraction of gravitino mass $m_{3/2}$. Scalar
              particles are generally much heavier than gauginos
              except for the limiting case of $p \rightarrow 1$.}}
        \label{fig:anomplot1}
    \end{center}
\end{figure}

Let us now investigate the pattern of soft terms as the parameters
$p_{ia}^{\alpha}$ and $p_{ij}^{\alpha}$ are varied by assuming that
$p_{ia}^{\alpha} = p_{ij}^{\alpha} \equiv p$
with $p$ a constant. If the scale at which the soft terms emerge is
taken to be $\Lambda_{\rm UV} = 1 \times 10 ^{11}$ GeV then the spectrum
of soft terms as a function of $p$ is displayed in
Figure~\ref{fig:anomplot1}. In general gaugino masses are an order of
magnitude smaller than scalar masses, except for values of $p$
approaching the limiting case of $p=1$ (which is equivalent to scenario
(B) given by~(\ref{B})) where scalar masses go through zero.  It
  is important to note that all of the possibilities of
  Figure~\ref{fig:anomplot1} represent ``anomaly mediated''
  scenarios. However, it is only the extreme case of $p=1$ that was
  studied previously in the particular model of Randall and
  Sundrum~\cite{rs}.

One final aspect of these soft term patterns relevant to low energy
phenomenology is the relative size of the scalar masses and
A-terms. It is well known that for any generation of matter with
non-negligible Yukawa couplings the relation
\begin{equation}
|A_{ijk}|^2 \leq 3(M_{i}^{2} + M_{j}^{2} + M_{k}^{2}),
\label{CCB}
\end{equation}
evaluated at the scale of supersymmetry breaking, is a good
indicator that the minimum of the scalar potential will yield proper
electroweak symmetry breaking: when the bound is not satisfied it is
typical to develop minima away from the electroweak symmetry breaking
point in a direction in which one of the scalars masses of a field
carrying electric or color charge becomes negative. Since the
``anomaly mediated'' A-term and the scalar mass {\em squared} both
have a single loop factor of $1/16\pi^{2}$ the condition~(\ref{CCB})
is generally satisfied. For example, in scenario (A) discussed above
\begin{equation}
(M_{i}^{2} + M_{j}^{2} + M_{k}^{2})= m_{3/2}A_{ijk},
\label{CCBscenA}
\end{equation}
and since $A_{ijk}$ is loop-suppressed relative to the gravitino mass,
as seen from~(\ref{anomAterm}), this scenario is phenomenologically
acceptable. Scenario (B) with its vanishing scalar masses at one loop
is problematic, however, and the two loop contributions are relevant
to the determination of its viability.

\subsection{The O-II models}
\label{sec:O2case}

\begin{figure}[thb]
    \begin{center}
\centerline{
       \epsfig{file=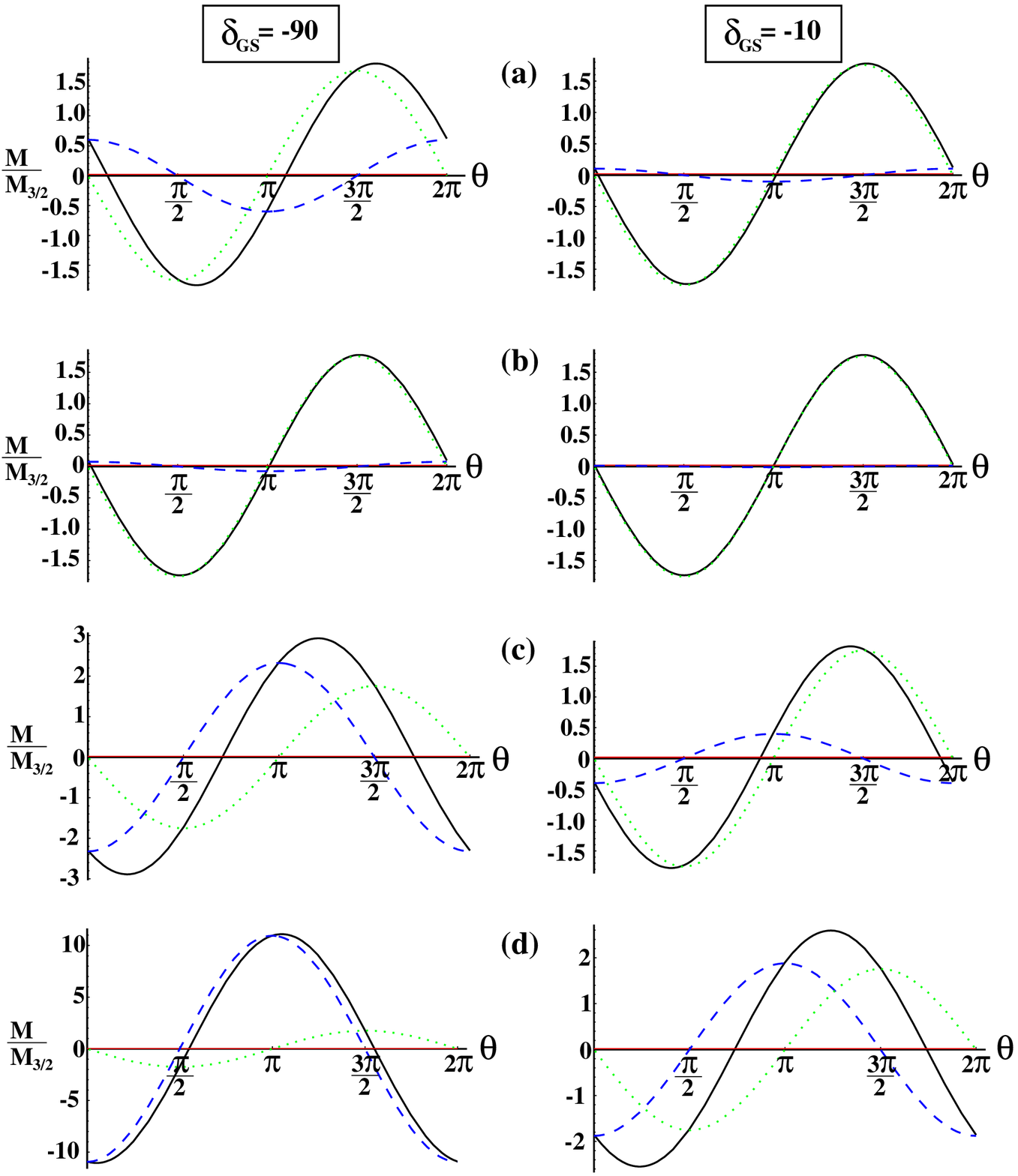,width=0.8\textwidth}}
          \caption{{\footnotesize {\bf $U(1)_{Y}$ Gaugino Mass in the
                BIM O-II Model}. Contributions to the value of
              $M_{1}$ from $F^{T}$ (dashed) and $F^{S}$ (dotted) as
              well as total $M_{1}$ (solid) are
              given as a function of Goldstino angle for two values of
              $\delta_{\rm GS}$ and four T-modulus {\it vevs}: $\lang
              {\rm Re}\;t \rang =0.5$ (a), $\lang {\rm Re}\;t \rang = 0.9$
              (b), $\lang {\rm Re}\;t \rang =5$ (c), and $\lang {\rm
                Re}\;t \rang= 20$ (d). All values are given as a
              fraction of the gravitino mass $m_{3/2}$.}}
        \label{fig:bimO2plot1}
    \end{center}
\end{figure}

This class of orbifold models discussed in \cite{bim} has matter fields in the untwisted
sector with weights $(n_i^1,n_i^2,n_i^3)=(-1,0,0)$ $(0,-1,0)$ or $(0,0,-1)$.
Then, taking for simplicity the same common value $T$ for the $T^\alpha$
fields\footnote{All the expressions given in this and the following
  sections will assume zero phases $\gamma_{S}=\gamma_{T}=0$}, one
obtains from~(\ref{MaFull}):
\begin{equation}
M_{a}^{tot}=\frac{g_{a}^{2}(\mu)}{2\sqrt{3}}\overline{M}\lbr
\frac{2}{\sqrt{3}}\cos\theta (t+
\overline{t}) G_{2} \[ \frac{\delta_{\rm GS}}{16\pi^{2}} + b_{a}^{0}\]
+\frac{2 b_{a}^{0}}{\sqrt{3}} +
\frac{\sin\theta}{k_{s\overline{s}}^{1/2}}\[ 1+\frac{g_{s}^{2}}{16\pi^2}\(C_a
-\sum_{i}C_{a}^{i}\)\]  \rbr.
\label{MaO2}
\end{equation}
The above form suggests a closer investigation of the relative
magnitude of the contributions to gaugino masses arising from the
dilaton sector (proportional to $\sin\theta$), the moduli sector
(proportional to $\cos\theta$) and the anomaly-induced piece
(independent of the Goldstino angle). As mentioned in the previous
section, any tree level contribution (from the
dilaton sector) will likely dominate the gaugino mass, particularly
when the Green-Schwarz coefficient is smaller than $-3C_{E_{8}}$. The
anomaly-induced piece is typically quite small and will only be
relevant in the case of moduli
domination ($\sin\theta=0$) with moduli stabilized very near their
self-dual points and/or very small Green-Schwarz coefficient. This
behavior is demonstrated for the case of the $U(1)_{Y}$ gaugino
mass $M_{1}$ in Figure~\ref{fig:bimO2plot1}. We have taken
$k=-\ln(S+\overline{S})$ and set $g_{s}^{2} =1/2$. 

\begin{figure}[t]
    \begin{center}
\centerline{
       \epsfig{file=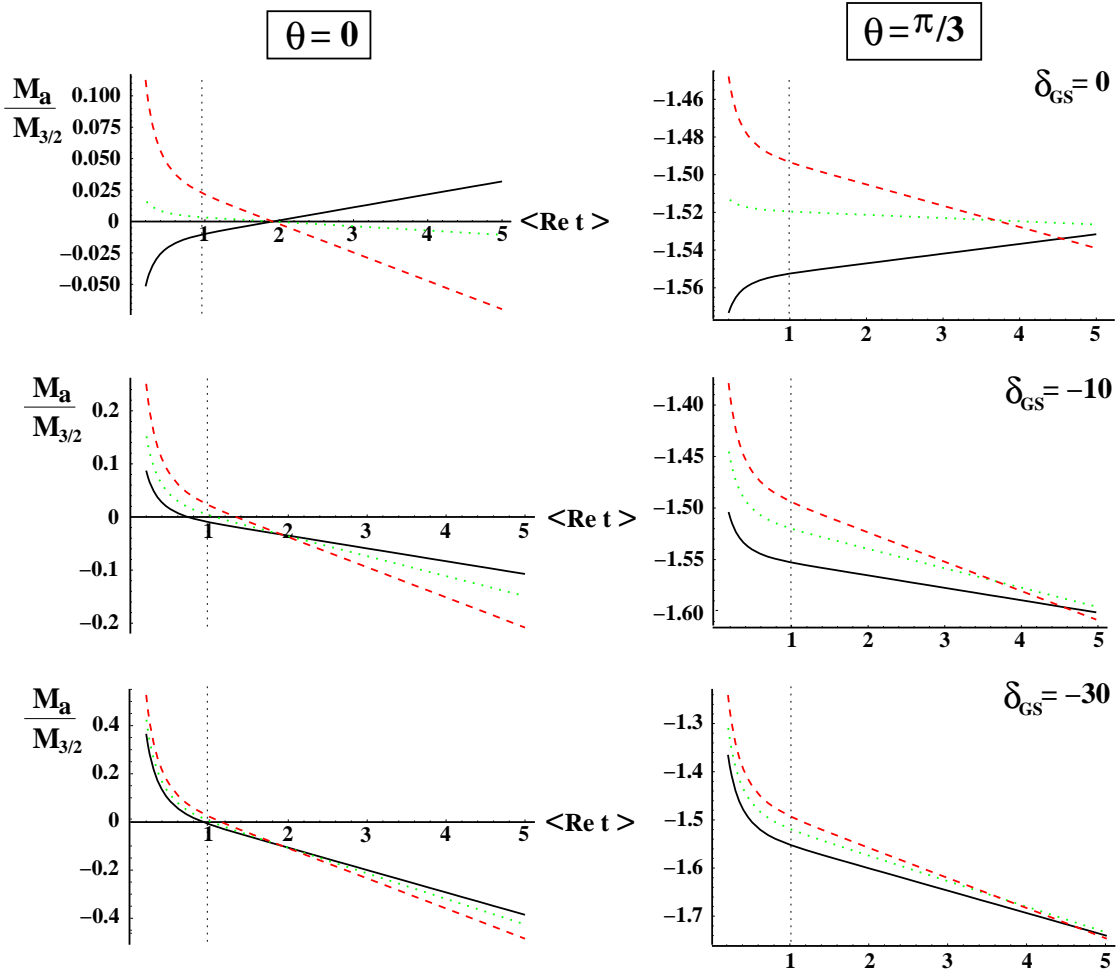,width=0.9\textwidth}}
          \caption{{\footnotesize {\bf Relative Gaugino Masses
                vs. $\lang {\rm Re}\;t \rang$ in the
                BIM O-II Model with $\Lambda_{\rm UV} = 2 \times
                10^{16}$ GeV}. Relative sizes of the three gaugino
              masses $M_{1}$ (dashed), $M_{2}$ (dotted) and $M_{3}$
              (solid) are displayed as a function of $\lang {\rm Re}\;t
              \rang$ for two values of the Goldstino angle $\theta$
              and three representative values of $\delta_{\rm
                GS}$. The vertical dotted line at $\lang {\rm Re}\;t
              \rang =1$ indicates the moduli self-dual point where
              gaugino masses become independent of $\delta_{\rm
                GS}$. When $\sin\theta=0$ this point represents the
              case of leading anomaly contributions discussed in
              Section~\ref{sec:anomalycase}. All masses are relative
              to the gravitino mass $m_{3/2}$.}}
       \label{fig:bimO2plot2}
    \end{center}
\end{figure}

In Figure~\ref{fig:bimO2plot2} we look at the
relative sizes of the three gaugino mass terms for the case of moduli
domination ($\theta = 0$) and a mixed case ($\theta = \pi/3$) for real
moduli vacuum values $\lang {\rm Re}\;t \rang$ at a boundary scale of
$\Lambda_{\rm UV} = 2 \times 10^{16}$ GeV. Note
that there is always a particular value of the moduli {\em vev} such
that a nearly degenerate gaugino mass spectrum is recovered. As
$\cos\theta \rightarrow 0$ this value gets ever larger as we approach
the limiting case in which the gaugino masses are independent of the
value of $\lang {\rm Re}\ t \rang$. At the GUT scale where $g_{2}^{2}
\approx g_{1}^{2} \approx 1/2$ the difference in SU(2) and U(1) gaugino masses is
given by
\begin{equation}
M_{2}-M_{1} \approx -\frac{m_{3/2}}{40\pi^{2}} \lbr 7\[1+\cos\theta
\(1-\frac{\pi}{3}{\rm Re}\ t\)\] +2\sqrt{3}\sin\theta \rbr,
\label{anomDiff}
\end{equation}
where we have used the fact that for ${\rm Re}\;t > 1$, $\zeta (t) \approx
-\pi/12$. For $\theta=0$ equation~(\ref{anomDiff}) indicates
that $M_{1}=M_{2}$ at ${\rm Re}\;t \approx 6/\pi$ while for
$\theta=\pi/3$ this occurs when ${\rm Re}\;t \approx 3.7$.

When $\theta \not= 0$~(\ref{anomDiff}) implies that $|M_{2}| \geq
|M_{1}|$ (the gaugino masses in this regime are negative) whenever 
\begin{equation}
{\rm Re}\ t \leq
\frac{3}{\pi}\lbr \frac{2\sqrt{3}}{7}\tan\theta + \sec\theta +1 \rbr.
\label{crossGUT}
\end{equation} 
In the case where $\theta=0$ so that there is no tree level contribution
to gaugino masses we see from Figure~\ref{fig:bimO2plot2} that $|M_{1}|
\geq |M_{2}|$ in nearly all of the $\lang {\rm Re}\; t\rang$ parameter
space. This relationship between the boundary values of the SU(2) and U(1)
gaugino masses is crucial to the low energy phenomenology of the model
in that it determines whether the lightest neutralino is predominantly
bino-like, predominantly wino-like or a mixed state. Thus a lightest
neutralino with a significant wino content {\em need not necessarily
  imply that supersymmetry breaking is due to pure anomaly
  mediation}. We will return to this point when we investigate sample
spectra in the next section.

The scale at
which the soft masses emerge is particularly important: the largest
contributions to gaugino masses generically arise from the tree level
piece and the piece proportional to the Green-Schwarz coefficient
$\delta_{\rm GS}$. These terms cancel in~(\ref{anomDiff}), however,
when the difference is evaluated at the GUT scale. Thus the location
of the crossover point is independent of the choice of $\delta_{\rm
  GS}$ in Figure~\ref{fig:bimO2plot2}.

\begin{figure}[t]
    \begin{center}
\centerline{
       \epsfig{file=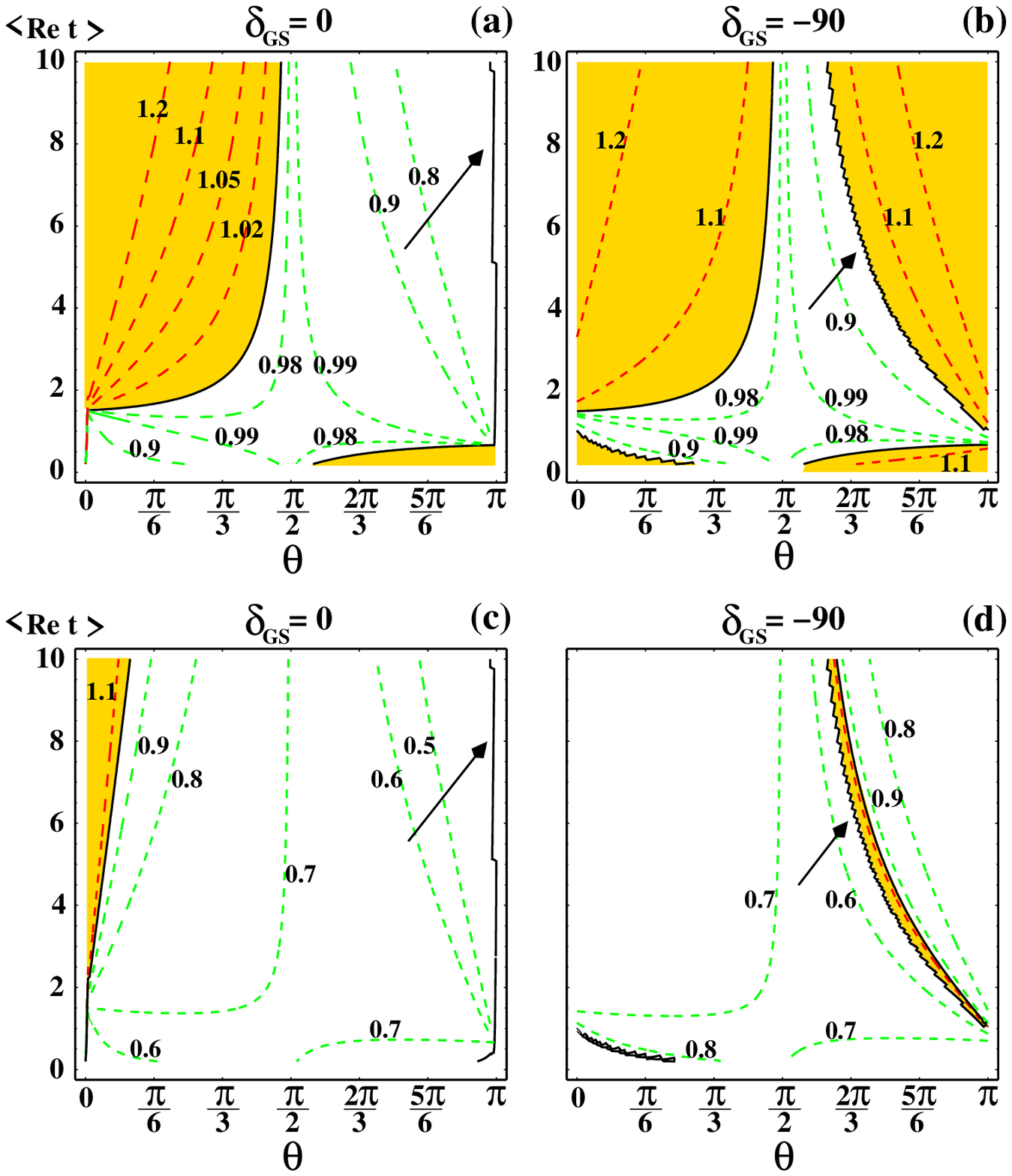,width=0.9\textwidth}}
          \caption{{\footnotesize {\bf Ratio $M_{1}/M_{2}$ in BIM O-II
                Model.} Contours of the absolute value of the ratio of
              U(1) to SU(2) gaugino
              masses are
              given for boundary scales of $\Lambda_{\rm UV} = 2 \times
                10^{16}$ GeV for panels (a) and (b), and $\Lambda_{\rm
                  UV} = 1 \times 10^{11}$ GeV for panels (c) and
                (d). The shaded area is the region of parameter space
                for which $|M_{1}| \geq |M_{2}|$. The arrow indicates the
                direction of smallest ratios as the discontinuity
                $M_{2} = 0$ is approached. The contour $M_{1} =
                M_{2}$ is given by the
                heavy solid line.}}
        \label{fig:contour}
    \end{center}
\end{figure}

An immediate consequence of the above is that measurement of the
properties of the lightest neutralinos may reveal information on the
nature of the scale of ultraviolet physics. In particular the region of
parameter space for which the lightest neutralino is predominantly
wino-like becomes increasingly small as the scale of
supersymmetry breaking is lowered. This is
illustrated in
Figure~\ref{fig:contour} where we plot the ratio of gaugino masses
$M_{1}/M_{2}$ for two different boundary scales: $\Lambda_{\rm UV}
= 2 \times 10^{16}$ GeV and $\Lambda_{\rm UV} = 1 \times 10^{11}$ GeV,
for which
$g_{2}^{2} \approx (7/5)g_{1}^{2}$. As the gauge couplings run farther
apart the shaded areas in which $M_{1}/M_{2} \geq 1$ (and hence where a
wino-like lightest neutralino is possible) grow steadily smaller. When
$\delta_{\rm GS} = 0$ the ratio $M_{1}/M_{2}$ diminishes as the
Goldstino angle $\theta$ increases until $M_{2}$ begins to approach
its vanishing value and the ratio passes through a discontinuity
before increasing rapidly as $\theta \rightarrow \pi$. When the
Green-Schwarz coefficient $\delta_{\rm GS}$ is increased the location
of this discontinuity, as indicated in Figure~\ref{fig:contour} by a
heavy arrow, moves to smaller values of $\theta$.

The trilinear A-terms for these orbifold models are given by~(\ref{Atermfull}):
\begin{eqnarray}
A^{tot}_{ijk} &=& \frac{\overline{M}}{\sqrt{3}} \lbr
-\frac{\gamma_{i}}{\sqrt{3}} -\frac{\cos\theta}{\sqrt{3}}
(t+\overline{t}) G_{2} \[ \sum_{a} \gamma_{i}^{a}
 p_{ia} + \sum_{lm} \gamma_{i}^{lm} p_{lm}\]  +
 \frac{\sin\theta}{k_{s\overline{s}}^{1/2}} 
   \[-\frac{k_{s}}{3} +\sum_{lm}
   \gamma_{i}^{lm} \partial_{s} \ln(\tmu^2_{lm}) \right. \right. \nonumber \\
 & & \left.  \left. +\sum_{a}\gamma_{i}^{a} \( \partial_{s}
   \ln(\tmu^2_{ia}) + \frac{g_{a}^{2}}{2}\ln(\tmu^2_{ia}) \)   -
     \ln\[(t+\overline{t}) |\eta(t)|^4\] \(
     \sum_{a} g_{a}^{2} \gamma_{i}^{a} p_{ia} - \sum_{lm} k_{s}
     \gamma_{i}^{lm} p_{lm} \) \] \rbr
\nonumber \\ & &     +\rm{cyclic}(ijk),  
\label{AtermO2}
\end{eqnarray}
where $p_{ia}=\sum_{\alpha} p_{ia}^{\alpha}$ and
$p_{lm}=\sum_{\alpha}p_{lm}^{\alpha}$. For scenario (A), as defined
by~(\ref{A}), this expression is particularly simple
\begin{equation}
A^{tot}_{ijk} = \frac{\overline{M}}{\sqrt{3}}\lbr \frac{\sin\theta}{
  k_{s\overline{s}}^{1/2}} \[-\frac{k_{s}}{3}
  + \sum_{a}\gamma_{i}^{a}
  \frac{g_{a}^{2}}{2}
  \ln(\tmu^2_{ia}/\mu_{R}^{2})\] -\frac{\gamma_{i}}{\sqrt{3}} \rbr
  +\rm{cyclic}(ijk).
\label{AtermO2A}
\end{equation}
It is worth noting that, with such a scenario for the PV metrics, this 
pattern for A-terms goes beyond the BIM O-II model.
Any of the following conditions: (i)
$\sum_{\alpha}(n^{\alpha}_{i} + n^{\alpha}_{j} + n^{\alpha}_{k} +1)=0$
with identical vacuum values for all T-moduli (as in the BIM O-II
model), (ii) $\cos\theta=0$ (dilaton domination) or (iii) $G_2^{\alpha}=0$
(moduli stabilized at self-dual point), yields the
A-terms given by~(\ref{AtermO2A}) above.

By contrast, for scenario (B) defined by~(\ref{B}) the A-terms take
the form
\begin{eqnarray}
A^{tot}_{ijk} &=& \frac{\overline{M}}{\sqrt{3}}\lbr
-\frac{\gamma_{i}}{\sqrt{3}}\[ 1 + \cos\theta (t +
\overline{t}) G_{2} \] +\frac{\sin\theta}{k_{s\overline{s}}^{1/2}}
\[ -\frac{k_{s}}{3} + \sum_{a} \frac{g_{a}^{2}}{2} \gamma_{i}^{a}
(\ln(g_{a}^{2})-1) \right. \right. \nonumber \\
 & & \left. \left.  -\ln\[(t + \overline{t})
     |\eta(t)|^4\] \( \sum_{a} g_{a}^{2}\gamma_{i}^{a} -
  \sum_{lm} k_{s} \gamma_{i}^{lm} \) \] \rbr +\rm{cyclic}(ijk).
\label{AtermO2B}
\end{eqnarray}
This scenario also allows for the recovery of an
``anomaly mediated-like'' result of A-terms proportional to anomalous
dimensions in the moduli dominated limit ($\sin\theta =
0$). Expression~(\ref{AtermO2B}) differs from the situation in
Section~\ref{sec:anomalycase} in that for moduli domination this
scenario can accommodate proper electroweak symmetry breaking provided
the moduli are stabilized {\em away} from their self-dual points: in
particular, using the fact that for ${\rm Re}\;t > 1$, $\zeta (t) \approx
-\pi/12$ we have $\lang (t+\overline{t})G_{2} \rang \approx -1$ for $\lang t
\rang \approx 6/\pi \approx 2$ leading to
$A \approx 0$ from~(\ref{AtermO2B}).

The expressions for the bilinear B-terms are similar, but with added
model dependence at the tree level involving the origin of bilinear
terms in the K\"ahler potential or superpotential. For the case of
scenario (A) the general form given in~(\ref{Btermfull}) yields
\begin{eqnarray}
B^{tot}_{ij}&=& \frac{\overline{M}}{\sqrt{3}}
 \frac{\sin\theta}{k_{s\overline{s}}^{1/2}} \[ -k_{s} -\partial_{s}
 \ln \mu_{ij} +\sum_{a} \gamma_{i}^{a} \(\frac{g_{a}^{2}}{2}\)
 \ln(\tmu^2_{ia}) \right.  \nonumber \\
 & & \left. + \frac{\overline{M}}{6} \cos\theta \[ 1 - \sum_{\alpha}
\partial_{t^{\alpha}} \ln \mu_{ij} \]
+\frac{\overline{M}}{3}\(\frac{1}{2} -\gamma_{i}\) \] + (i
   \leftrightarrow j),
\label{BtermO2A}
\end{eqnarray}
while for case (B) the corresponding expression is
\begin{eqnarray}
B^{tot}_{ij}&=& \frac{\overline{M}}{\sqrt{3}}
 \frac{\sin\theta}{k_{s\overline{s}}^{1/2}} \[ -k_{s} -\partial_{s}
 \ln \mu_{ij}  +\sum_{a} \gamma_{i}^{a} \frac{g_{a}^{2}}{2}\(
 \ln(g_{a}^{2})-1\)  -\ln\[(t + \overline{t})|\eta(t)|^4\] \( \sum_{a}
   \gamma_{i}^{a} g_{a}^{2} - \sum_{lm} \gamma_{i}^{lm} k_{s} \) \] \nonumber \\
 & &  +\frac{\overline{M}}{3}\(\frac{1}{2} -\gamma_{i}\)
+\frac{\overline{M}}{6} \cos\theta \[ 1 - \sum_{\alpha}
\partial_{t^{\alpha}} \ln \mu_{ij} -2 (t+\overline{t}) G_{2} 
\gamma_{i} \] + (i \leftrightarrow j).
\label{BtermO2B}
\end{eqnarray}

Finally, the scalar masses in the BIM O-II model are found
from equation~(\ref{massfull}) of Appendix~B. Under the assumptions of
scenario (A) this reduces to
\begin{eqnarray}
(M_{i}^{tot})^{2}&=&|M|^{2}\lbr\frac{1}{9}\gamma_{i} +
\frac{1}{k_{s\overline{s}}^{1/2}}
   \frac{\sin\theta}{3\sqrt{3}} \[ \sum_{a} \gamma_{i}^{a} g_{a}^{2} -
   \frac{1}{2} 
   \sum_{jk} \gamma_{i}^{jk} (k_{s} + k_{\overline{s}}) \]+
   \frac{\sin^{2}\theta}{9} \[1 - \sum_{a}\gamma_{i}^{a}
   \ln(\tmu^2_{ia})  \right. \right. \nonumber \\
 & & \left. \left. + 2\sum_{jk}
 \gamma_{i}^{jk} \ln(\tmu^2_{jk}) \] + \frac{\sin^{2}\theta}{k_{s\overline{s}}} \[ -\frac{1}{4}
    \sum_{a}  g_{a}^{4} \gamma_{i}^{a} \ln(\tmu^2_{ia}) -\frac{1}{3} \sum_{jk} \gamma_{i}^{jk}
    \(k_{s}k_{\overline{s}} +2k_{s\overline{s}}\)
    \ln(\tmu^2_{jk}) \] \rbr,
\label{massO2A}
\end{eqnarray}
and for scenario (B) the scalar masses are given by
\begin{eqnarray}
(M_{i}^{tot})^{2}&=&|M|^{2}\lbr \frac{1}{3\sqrt{3}}
\frac{\sin\theta}{k_{s\overline{s}}^{1/2}} \[ 1 + \cos\theta
(t+\overline{t}) G_{2} \] \[ \sum_{a} g_{a}^{2}
\gamma_{i}^{a} - \frac{1}{2} \sum_{jk} \gamma_{i}^{jk} \(k_{s} +
k_{\overline{s}}\) \] \right. \nonumber \\
 & & \left. + \frac{\sin^{2}\theta}{9} \[ 1 + \gamma_{i} + \ln\[(t +
   \overline{t})|\eta(t)|^4\] \( \sum_{a} \gamma_{i}^{a} -2 \sum_{jk}
   \gamma_{i}^{jk} \) - \sum_{a} \gamma_{i}^{a} \ln(g_{a}^{2}) +2 \sum_{jk}
   \gamma_{i}^{jk} \ln(\tmu^2_{jk}) \] \right. \nonumber \\
 & & \left. -\frac{\sin^{2}\theta}{k_{s\overline{s}}} \[ \sum_{a}
   \gamma_{i}^{a} \(\frac{g_{a}^{4}}{4}\) \(\ln(g_{a}^{2})
   +\frac{5}{3}\)
   +\frac{1}{3} \sum_{jk} \gamma_{i}^{jk} \(k_{s}k_{\overline{s}}
   +2k_{s\overline{s}}\) \ln(\tmu^2_{jk}) \right. \right. \nonumber \\
 & & \left. \left. 
+ \ln\[(t +
   \overline{t})|\eta(t)|^4\] \( \sum_{a} \gamma_{i}^{a}
   \(\frac{g_{a}^{4}}{4}\) +\frac{1}{3} \sum_{jk} \gamma_{i}^{jk}
   k_{s} k_{\overline{s}} \) \] \rbr.
\label{massO2B}
\end{eqnarray}

The pattern of soft supersymmetry breaking terms that arise in this
orbifold model with uniform modular weights $n_{i}=-1$ and with the
same K\"ahler metric for the $\Pi^A$ and the $\Phi^A$, as
in scenario~(\ref{B}), will produce a low energy phenomenology very similar to
that of the recently proposed ``gaugino mediation''
scenario~\cite{gauginomed} if the Green-Schwarz coefficient is
sufficiently large, the supersymmetry breaking is moduli dominated and
the moduli are stabilized at $\lang {\rm Re}\;t \rang
\approx 2$. Such a situation gives rise to exactly vanishing scalar
masses and nearly vanishing A-terms and the gaugino masses in such a regime are
very nearly universal, as can be seen from the lower panels of
Figure~\ref{fig:bimO2plot2}. However, as the Green-Schwarz coefficient
is reduced the gaugino masses become negligible at the point $\lang
{\rm Re}\;t \rang \approx 2$, eventually coming into conflict with
direct search results at LEP and the Tevatron. 
Specific spectra for the O-II model will be presented with spectra for
orbifold models with large threshold corrections, to which we now turn.

\subsection{The O-I models}
\label{sec:O1case}

Models of this type were proposed with the goal of obtaining
coupling constant unification at the string scale, as opposed to the
extrapolated unification scale of $\Lambda_{\rm GUT} \approx 2
\times 10^{16}$ GeV which is typically a factor of twenty or so lower
than the string scale. This is achieved through large string threshold
corrections and the requirement of both particular sets of modular
weights for the massless fields and relatively large values of $\lang
{\rm Re}\; t \rang$ far from the self-dual points. Other solutions to this
discrepancy of scales have been proposed since but because the O-I
models have often been discussed in the literature we include them in
the present discussion.

\begin{figure}[t]
    \begin{center}
\centerline{
       \epsfig{file=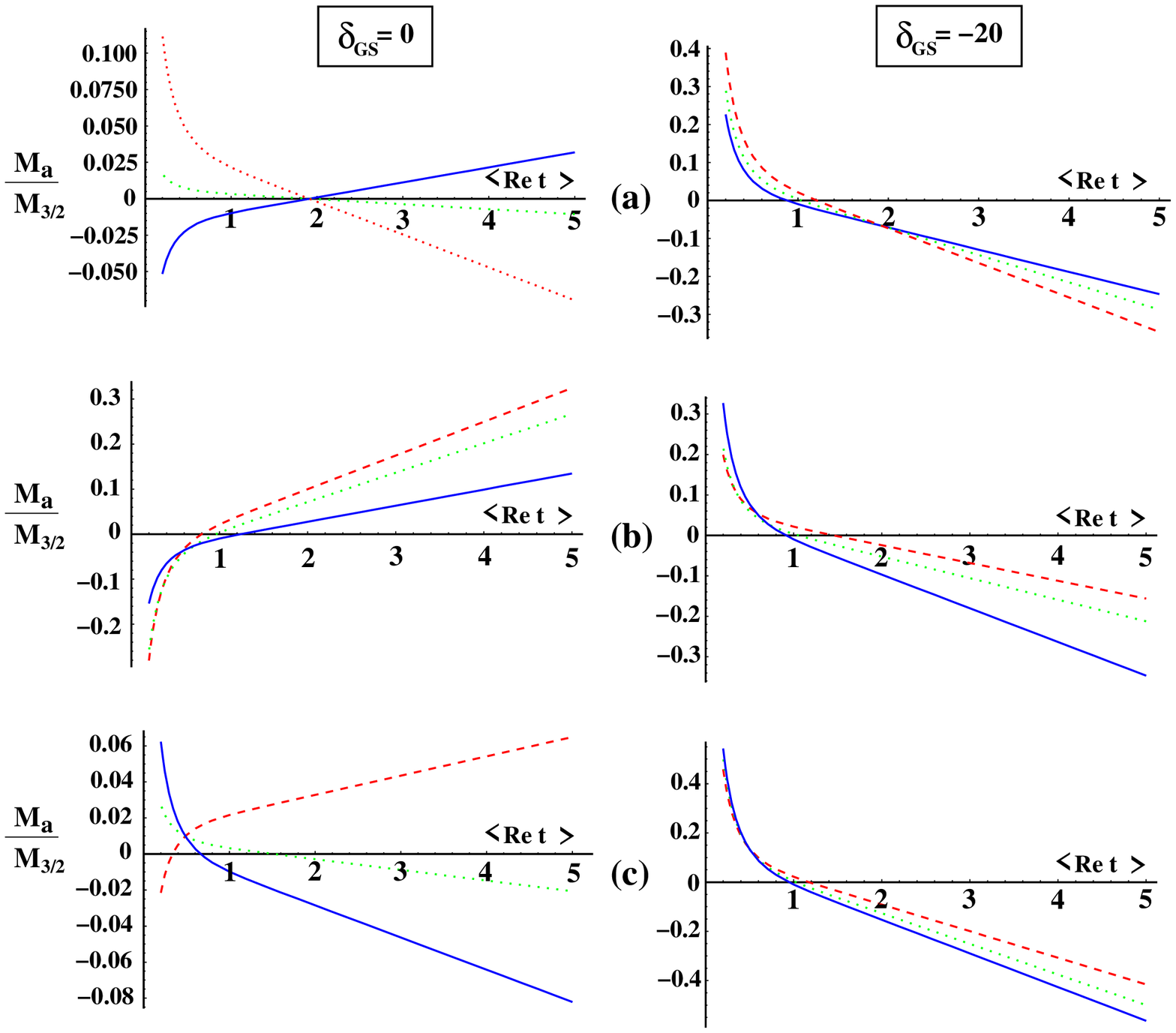,width=0.9\textwidth}}
          \caption{{\footnotesize {\bf Relative Gaugino Masses in the
                BIM O-II and BIM O-I Models with $\theta=0$}.
              Relative sizes of the three gaugino
              masses $M_{1}$ (dashed), $M_{2}$ (dotted) and $M_{3}$
              (solid) are displayed as a function of $\lang {\rm Re}\;t
              \rang$ for two values of the Green-Schwarz coefficient
              $\delta_{\rm GS}$ and $\Lambda_{\rm UV} = 2 \times
                10^{16}$ GeV. The top panels (a) represent the BIM
              O-II model from Section~\ref{sec:O2case}, the middle
              panels (b) represent the BIM O-I model and the bottom
              panels (c) represent the Love \& Stadler case
              from~\cite{Love}. All
              masses are relative
              to the gravitino mass $m_{3/2}$.}}
        \label{fig:bimO1plot1}
    \end{center}
\end{figure}

To investigate the phenomenological consequences of such models we
will assume a common
vacuum value for all three moduli and take $\Theta_{\alpha} =
1/\sqrt{3}$ as before. We shall investigate
two scenarios: (A) the original
``O-I'' scenario of
Brignole et al.~\cite{bim} with modular weights $n_{Q}=n_{D}=-1$, $n_{U}=-2$,
$n_{L}=n_{E}=-3$, $n_{H_d}, n_{H_u} = -4$ and (B) a $Z_{3} \times Z_{6}$
compactification studied by Love and Stadler~\cite{Love} with modular
weights $n_{Q}=n_{D}=0$, $n_{U}=-2$,
$n_{L}=-4$, $n_{E}=-1$, $n_{H_d}=n_{H_u} = -1$. In what follows let us
assume that the soft terms emerge at a scale for which logarithms such
as $\ln(\tmu^2_{ia})$ and $\ln(\tmu^2_{jk})$ are negligible and assume
PV case (A) for simplicity. In this approximation the general
expressions of
Appendix~B take a simplified form. The gaugino masses, given by
\begin{eqnarray}
M_{a}^{tot}&=&g_{a}^{2}(\mu)\frac{\overline{M}}{3}\lbr
b_{a}^{0} + \cos\theta (t+
\overline{t}) G_{2} \[ \frac{\delta_{\rm GS}}{16\pi^{2}}
+ b_{a}^{0}-\frac{1}{8\pi^{2}}\sum_{i}C_{a}^{i}(1+n_{i})\] 
 \right. \nonumber \\
 & & \left.
+
\frac{\sqrt{3}\sin\theta}{2k_{s\overline{s}}^{1/2}}\[ 1+\frac{g_{s}^{2}}{16\pi^2}\(C_a
-\sum_{i}C_{a}^{i}\)\] \rbr,
\label{MaO1}
\end{eqnarray}
are displayed in
Figure~\ref{fig:bimO1plot1} with the
value $\theta = 0$ (where the impact of the differing modular weights
is the greatest) for three models:
the BIM O-II case of Section~\ref{sec:O2case}, the original
BIM O-I case and the Love \& Stadler case.  The boundary scale is taken
to be $\Lambda_{\rm UV} = 2 \times 10^{16}$ GeV.\footnote{Though these
models are designed to allow for unification of gauge couplings at the
string scale $\Lambda_{\rm str} \approx 5 \times 10^{17}$ GeV, we will
investigate the pattern of soft supersymmetry-breaking terms at the GUT
scale to allow for comparison with other models.} It is clear from
Figure~\ref{fig:bimO1plot1} that the
modular weights of the matter fields play a crucial role in
determining the gaugino mass spectrum  provided the
Green-Schwarz coefficient is sufficiently small. As this parameter is
increased it will quickly come to dominate the other terms in~(\ref{MaO1}).

\begin{figure}[t]
    \begin{center}
\centerline{
       \epsfig{file=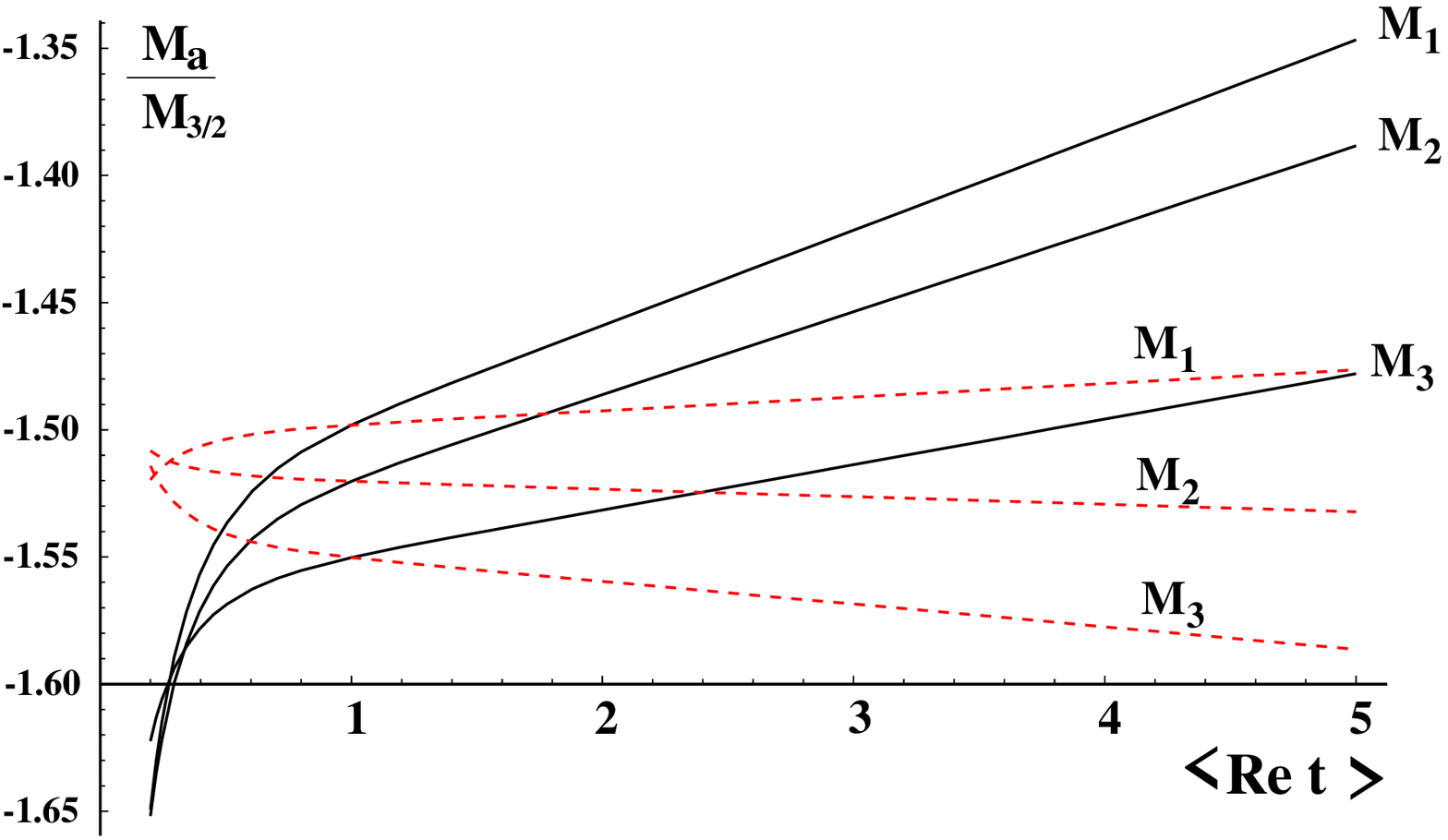,width=0.7\textwidth}}
          \caption{{\footnotesize {\bf Relative Gaugino Masses in the
                BIM O-I Models with $\theta =\pi/3$ and $\delta_{\rm GS}=0$}.
              Relative sizes of the three gaugino
              masses are displayed as a function of $\lang {\rm Re}\;t
              \rang$ for the BIM O-I Model (solid) and the Love \&
              Stadler Model (dashed) at $\Lambda_{\rm UV} = 2 \times
                10^{16}$ GeV. All
              masses are relative
              to the gravitino mass $m_{3/2}$.}}
        \label{fig:bimO1plot2}
    \end{center}
\end{figure}

However, looking at the tree level expressions for the scalar
masses~(\ref{treeterms}) it is apparent that when $\cos\theta =1$ any
field with a modular weight such that $n_{i} < -1$ will have a
negative tree level scalar mass-squared, as was noted in~\cite{bim}. 
Thus, to accommodate these large threshold models proper electroweak
symmetry breaking ({\it i.e.} positive scalar mass-squareds) will
generally require a Goldstino angle such that
$\sin\theta$ is large and the tree level terms in~(\ref{MaO1}) are
dominant. Models with a viable low energy
vacuum will therefore be models for which the impact of the matter fields'
modular weights on the gaugino spectrum is considerably muted. This is
displayed in Figure~\ref{fig:bimO1plot2} where
gaugino masses in the
BIM O-I model and the Love \& Stadler model are displayed for $\theta =
\pi/3$ and $\delta_{\rm GS} = 0$. We see that in these realistic cases
the differences in gaugino mass spectra between these models is small,
making them hard to distinguish experimentally.

The trilinear A-terms for scenario (A) are
\begin{equation}
A^{tot}_{ijk} = \frac{\overline{M}}{3} \lbr
-\gamma_{i} +\frac{\cos\theta}{3}
(t+\overline{t}) G_{2}  (n_{i}+ n_{j} + n_{k} +3) -
\frac{\sin\theta}{\sqrt{3}} \frac{k_{s}}{k_{s\overline{s}}^{1/2}}
    \rbr +\rm{cyclic}(ijk),  
\label{AtermO1A}
\end{equation}
and the scalar masses are determined from
\begin{eqnarray}
(M_{i}^{tot})^{2}&=&\frac{|M|^{2}}{9}\lbr (1+\gamma_{i}) +
   \cos\theta (t+\overline{t}) G_{2} \sum_{jk}
   \gamma_{i}^{jk} (n_{i}+ n_{j} + n_{k} +3) +n_{i}\cos^{2}\theta
 \right. \nonumber \\
 & & \left. +\frac{\sqrt{3}\sin\theta}{k_{s\overline{s}}^{1/2}}
   \( \sum_{a} \gamma_{i}^{a} g_{a}^{2} - \frac{1}{2}
   \sum_{jk} \gamma_{i}^{jk} \(k_{s} +
   k_{\overline{s}}\) \) \rbr,
\label{massO1A}
\end{eqnarray}
With these expressions we are now in a position to compare the typical
spectra of these O-I large threshold models with the models of
Section~\ref{sec:anomalycase} and Section~\ref{sec:O2case}.

\begin{table}[thb]
{\begin{footnotesize} {\begin{center}
\begin{tabular}{|l|c|ccccc|c|c|} \cline{1-9}
Model  & Anomaly~(\ref{sec:anomalycase}) &
\multicolumn{5}{c}{BIM O-II~(\ref{sec:O2case})} & BIM
O-I~(\ref{sec:O1case}) & L\&S~(\ref{sec:O1case})~\cite{Love} \\ \cline{1-9}
$\theta$ & 0 & 0 & 0 & 0 & 0 & $\pi/3$ & $\pi/3$ & $\pi/3$ \\
$\delta_{\rm GS}$ & N/A & 0 & 0 & 0 & -90 & -90 & -90 & -90 \\
$\lang {\rm Re}\;t \rang$ & 1 & $6/\pi$ & 5 & 20 & $6/\pi$ &
 $6/\pi$ & 16 & 14.5 \\
\cline{1-9}
$m_{3/2}$ & $1.9 \times 10^{4}$ & $1.9 \times 10^{4}$ & $1.6 \times
10^{4}$ & 4500 & 1600 & 450 & 150 & 150 \\
$m_{\tilde{N_1}}$ & 51.81 & 0.32 & 152.53 & 248.97 & 332 & 313 & 287 & 297 \\
$m_{\tilde{N_2}}$ & 168 & 3.7 & 462 & 759 & 615 & 599 & 557 & 581 \\
$\tilde{B}$ \% & 0.01 & 80.9 & 0.001 & 0.001 & 99.9 & 99.9 & 99.9 & 99.9 \\
$\tilde{W_{3}}$  \% & 99.7 & 19.1 & 99.7 & 99.7 & 0.001 & 0.001 & 0.001 & 0.001 \\
$m_{\tilde{\chi_{1}}^{\pm}}$ & 51.83 & 3.1 & 152.55 & 249.00 & 615 & 599 & 557 & 581 \\
$m_{\tilde{g}}$ & 623 & 3.6 & 1468 & 2245 & 2156 & 2164 & 2106 & 2128 \\
$m_{h}$ & 114 & 114 & 114 & 114 & 114 & 114 & 114 & 114 \\
$m_{A}$ & 2237 & 2217 & 1992 & 1357 & 1447 & 1387 & 1810 & 1568 \\
$m_{\tilde{t}_{R}}$ & 860 & 796 & 1142 & 1597 & 1521 & 1610 & 1373 & 1532 \\
$m_{\tilde{t}_{L}}$ & 1842 & 1810 & 1818 & 1820 & 1804 & 1866 & 1709 & 1793 \\
$m_{\tilde{b}_{R}}$ & 1805 & 1765 & 1802 & 1769 & 1782 & 1847 & 1701 & 1773 \\
$m_{\tilde{b}_{L}}$ & 1810 & 1770 & 1824 & 1908 & 1883 & 1945 & 1881 & 1871 \\
$m_{\tilde{\tau}_{R}}$ & 1191 & 1180 & 1076 & 514 & 329 & 302 & 198 & 290 \\
$m_{\tilde{\tau}_{L}}$ & 1193 & 1182 & 1078 & 515 & 330 & 303 & 281 & 301 \\
$A_{\rm top}$ & 391 & 71 & -815 & -1423 & 1696 & 1541 & 560 & 1607 \\
$A_{\rm bot}$ & 973 & 463 & -999 & -1827 & 2819 & 2200 & -405 & 4650 \\
$A_{\rm tau}$ & 220 & 273 & 376 & 305 & 466 & -184 & -7417 & -2734 \\
$\mu$ & 1617 & 1592 & 1501 & 1281 & 1341 & 1302 & 1577 & 1297 \\
\cline{1-9}
\end{tabular}
\end{center}} 
\end{footnotesize}}
{\footnotesize \caption{{\bf Sample Spectra (in GeV) for Typical Models of
    Sections~\ref{sec:anomalycase}, \ref{sec:O2case}
    and~\ref{sec:O1case}.}  All cases are for PV scenario (A),
  $\tan\beta=3$ and $\Lambda_{\rm UV} = 2 \times
  10^{16}$ GeV ($\tilde{B}$ \%  and $\tilde{W_3}$ \% represent the
    content of the lightest neutralino in per cents). The first O-II
    case considered, while clearly ruled out experimentally, is
    presented as an illustrative example.}} 
\label{tbl:spectra}
\end{table}

In
Tables~2 and~3 we give some
representative sample spectra
for  Pauli-Villars scenario (A) defined
by~(\ref{A}) and $\tan\beta =3$ and $\tan\beta=10$, respectively. The
spectra for scenario (B) are very similar and these
values vary only minimally when $\Lambda_{\rm UV}$ is varied. To
obtain these spectra at the
electroweak scale the renormalization group equations (RGEs) were run
from the boundary scale to the electroweak scale. All gauge and Yukawa
couplings as well as gaugino masses and A-terms were run with one loop
RGEs while scalar masses were run at two loops to capture the possible
effects of heavy scalars on the evolution of third generation squarks
and sleptons. We chose to keep only the top, bottom and tau Yukawas
and the corresponding A-terms. The gravitino mass has been scaled in each
case to obtain a Higgs mass of 114 GeV, which
can be considered either as a limiting case or as an experimental
requirement, depending on what happens next at LEP.

At the electroweak scale the one loop corrected effective
potential $V_{\rm 1-loop}=V_{\rm tree} + \Delta V_{\rm rad}$ is
computed and the effective $\mu$-term $\bar{\mu}$ is calculated
\begin{equation}
{\bar{\mu}}^{2}=\frac{\(m_{H_d}^{2}+\delta m_{H_d}^{2}\) -
  \(m_{H_u}^{2}+\delta m_{H_u}^{2}\) \tan{\beta}}{\tan^{2}{\beta}-1}
-\frac{1}{2} M_{Z}^{2}.
\label{eq:radmuterm}
\end{equation}
In equation~(\ref{eq:radmuterm}) the quantities $\delta m_{H_u}$ and $\delta m_{H_d}$ are
the second derivatives of the radiative corrections $ \Delta V_{\rm
  rad}$ with respect to the up-type and down-type Higgs scalar fields, 
respectively. These corrections include the effects of all
third generation particles. If the right hand side of equation~(\ref{eq:radmuterm})
is positive then there exists some initial value of $\mu$ at the
condensation {\mbox scale} which results in correct electroweak symmetry
breaking with $M_{Z} = 91.187$~GeV.\footnote{Note that
  for these tables we do not 
try to specify the origin of this $\mu$-term (nor its associated B-term)
and merely leave them as free parameters in the theory --
ultimately determined by the requirement that the Z-boson receive the
correct mass.} 

Note that the gravitino mass varies greatly over the models considered
in Tables~2 and~3. For the anomaly case (which is equivalent
to the BIM O-II model with $\sin\theta=0$ and  $\lang {\rm Re}\;t \rang
= 1$) there is a large hierarchy between scalars and gauginos, as
noted in Section~\ref{sec:anomalycase}, which necessitates a large
value of the gravitino mass to yield neutralinos with masses near the
current LEP limits. Having normalized our scales to yield Higgs masses
of 114 GeV we find chargino masses (for PV scenario (A) and thus $p=0$
in Figure~\ref{fig:anomplot1}) below the recently reported bounds
of $m_{\chi^{\pm}} \geq 86.1$ GeV
for the case of a chargino which is nearly degenerate with a wino-like
lightest neutralino~\cite{L3}. As the PV scenario assumed is
modified, however, this relation between the chargino mass and Higgs
mass varies. In particular as the value of $p$ approaches larger,
positive values the gauginos steadily become heavier for a fixed Higgs
mass, eventually satisfying the experimental constraints. For the
large threshold models, by contrast, the
large values of $\lang {\rm Re}\;t \rang$ necessary to ensure gauge
coupling unification at the string scale make the gauginos typically
{\em heavier} than the gravitino at the boundary scale $\Lambda_{\rm
  UV}$, due to the large value of $(t+\overline{t}) G_{2}$, and have a
smaller degree of hierarchy between gauginos and scalars.

\begin{table}[thb]
{\begin{footnotesize} {\begin{center}
\begin{tabular}{|l|c|ccccc|c|c|} \cline{1-9}
Model  & Anomaly~(\ref{sec:anomalycase}) &
\multicolumn{5}{c}{BIM O-II~(\ref{sec:O2case})} & BIM
O-I~(\ref{sec:O1case}) & L\&S~(\ref{sec:O1case})~\cite{Love} \\ \cline{1-9}
$\theta$ & 0 & 0 & 0 & 0 & 0 & $\pi/3$ & $\pi/3$ & $\pi/3$ \\
$\delta_{\rm GS}$ & N/A & 0 & 0 & 0 & -90 & -90 & -90 & -90 \\
$\lang {\rm Re}\;t \rang$ & 1 & $6/\pi$ & 5 & 20 & $6/\pi$ &
 $6/\pi$ & 16 & 14.5 \\
\cline{1-9}
$m_{3/2}$ & 8000 & 8000 & 6500 & 1800 & 1200 & 200 & N/A & N/A \\
$m_{\tilde{N_1}}$ & 20.20 & 0.17 & 62.11 & 98.72 & 139 & 129 & & \\
$m_{\tilde{N_2}}$ & 70 & 3.11 & 187 & 301 & 260 & 244 & & \\
$\tilde{B}$ \% & 0.08 & 79.2 & 0.002 & $1.9 \times 10^{-7}$ & 99.3 & 99.1 & & \\
$\tilde{W_{3}}$ \% & 98.0 & 20.8 & 97.8 & 97.4 & 0.001 & 0.002 & & \\
$m_{\tilde{\chi_{1}}^{\pm}}$ & 20.21 & 2.5 & 62.14 & 98.75 & 260 & 244 & & \\
$m_{\tilde{g}}$ & 280 & 1.85 & 644 & 978 & 1020 & 979 & & \\
$m_{h}$ & 114 & 114 & 114 & 114 & 114 & 114 & & \\
$m_{A}$ & 797 & 790 & 689 & 485 & 560 & 497 & & \\
$m_{\tilde{t}_{R}}$ & 449 & 427 & 527 & 658 & 663 & 667 & & \\
$m_{\tilde{t}_{L}}$ & 797 & 782 & 774 & 806 & 849 & 819 & & \\
$m_{\tilde{b}_{R}}$ & 739 & 720 & 727 & 737 & 792 & 771 & & \\
$m_{\tilde{b}_{L}}$ & 763 & 744 & 753 & 799 & 838 & 812 & & \\
$m_{\tilde{\tau}_{R}}$ & 493 & 490 & 431 & 206 & 147 & 121 & & \\
$m_{\tilde{\tau}_{L}}$ & 503 & 499 & 440 & 211 & 156 & 132 & & \\
$A_{\rm top}$ & 190 & 47 & -336 & -596 & 796 & 668 & & \\
$A_{\rm bot}$ & 398 & 187 & -403 & -858 & 1223 & 893 & & \\
$A_{\rm tau}$ & 83 & 108 & 153 & 130 & 190 & 100 & & \\
$\mu$ & 578 & 565 & 529 & 495 & 559 & 499 & & \\
\cline{1-9}
\end{tabular}
\end{center}} 
\end{footnotesize}}
{\footnotesize \caption{{\bf Sample Spectra (in GeV) for Typical Models of
    Sections~\ref{sec:anomalycase}, \ref{sec:O2case}
    and~\ref{sec:O1case}.}  The same as in Table~2 but for
  $\tan\beta=10$. Neither of the
    large threshold models are viable at this value of $\tan\beta$.}} 
\label{tbl:spectra2}
\end{table}

The O-II
models can interpolate between these two extremes. When $\theta=0$ and
$\delta_{\rm GS}=0$ the pattern of physical masses shows the anomaly
mediated feature of a wino-like LSP. As the value of $\lang {\rm
  Re}\;t \rang$ increases from $\lang {\rm
  Re}\;t \rang =1$ (the pure anomaly mediated case) it first passes
through the experimentally excluded values where $\lang {\rm
  Re}\;t \rang \approx 6/\pi$ and the gaugino masses are nearly
zero. Thereafter the hierarchy between gauginos and scalars steadily
decreases until the spectra of masses is very similar to that of the
more typical supergravity spectra to the right of Table~2. However, as
mentioned at the end of the previous section the feature of a
wino-like LSP persists. Once $\theta \not= 0$ and/or $\delta_{\rm GS}
\not= 0$ the pattern of soft terms immediately becomes relatively
insensitive to the value of $\lang {\rm Re}\;t \rang$ and the LSP once
again becomes predominantly bino-like.

The models with large threshold corrections also tend to have very
light staus. In fact, as the value of $\tan\beta$ increases the stau
mass $m_{\tilde{\tau}_{R}}$ eventually becomes negative. The limiting value of
$\tan\beta$ for which these models are phenomenologically viable
depends slightly on the value of $\delta_{\rm GS}$: for $\theta =
\pi/3$ the model of Love
\& Stadler requires $\tan\beta < 9.1$ when $\delta_{\rm GS}=-90$ and
$\tan\beta < 4.8$ when $\delta_{\rm GS}=0$, while the original BIM O-I
model requires $\tan\beta < 3.1$ when $\delta_{\rm GS}=-90$ and is not
allowed at all for $\delta_{\rm GS}=0$. This is reflected in the empty
columns in Table~3. This problem is slightly ameliorated when the
Goldstino angle is increased. For $\theta = 2\pi/5$, for example,  the
model of Love
\& Stadler requires $\tan\beta < 12.7$ when $\delta_{\rm GS}=-90$ and
$\tan\beta < 9.6$ when $\delta_{\rm GS}=0$, while the original BIM O-I
model requires $\tan\beta < 4.9$ when $\delta_{\rm GS}=-90$ and
$\tan\beta < 2.1$ when $\delta_{\rm GS}=0$. 

The pattern of masses exhibited in Tables~2 and~3 suggests
  that the hierarchy between gauginos and scalars in any potential
  observation of supersymmetry will be a key to understanding the
  nature of the underlying physics giving rise to supersymmetry
  breaking. The observation of a lightest neutralino with significant
  wino content will not be enough to distinguish between the pure
  anomaly mediated cases and the BIM O-II type models but {\em will}
  indicate that supersymmetry breaking is moduli dominated within this
  class of models. The presence of
  a large hierarchy between scalars and gauginos and large mixing in
  the stop sector will point towards moduli stabilized at or near
  their self-dual values, while the absence of such effects would
  suggest the moduli are stabilized far from these values.

%The case of pure anomaly-mediation is the simplest to
%  distinguish: an LSP which is nearly pure wino with an order of
%  magnitude difference in the scale of gaugino masses and scalar
%  masses. If the lightest neutralino is instead mostly or completely
%  bino-like, then the absence of a hierarchy between gauginos and
%  scalars favors a scenario in which moduli are stabilized at or near
%  their self-dual points and modular weights are small. The presence
%  of a large degree of non-universality in the scalar mass spectrum
%  and a moderate hierarchy between scalars and gauginos will likely be
%  the result of large moduli values and/or large and non-universal
%  modular weights for the observable sector fields.
%{\em *** This
%  would not have been noticed by BIM in their papers since they didn't
%  have one loop corrections in a modular-invariant formalism. I think
%  a gravitino just slightly less massive than the gauginos is
%  problematic -- yet another nail in the coffin of these models? ***}

\subsection{The BGW model}
\label{BGWcase}

In this section we give the soft supersymmetry breaking parameters for
the model of Ref.~\cite{bgw}, with an explicit mechanism for
supersymmetry breaking through gaugino condensation in a hidden
sector, and dilaton stabilization by nonperturbative string effects.
An effective Lagrangian below the scale $\mu_c$ of hidden gaugino
condensation is constructed~\cite{BDQQ,BGT} by replacing the linear
multiplet $L$ in (\ref{GS}) by a vector multiplet $V$ whose components
includes those of $L$ and of a chiral multiplet $U$ and its conjugate
$\U$.  The superfield $U$ satisfies the same equations as the
composite chiral superfield $\hat U = W^\alpha W_\alpha$ constructed
from the Yang-Mills superfield strength, and is interpreted as the
lightest chiral superfield bound state of the effective theory below
the condensation scale $\mu_c = |u|^{1\over3},$ where $u = U|$ is the
scalar component of the chiral supermultiplet $U$.  An effective
potential for the gaugino condensates $U$, as well as matter
condensates $\Pi$ that are present if there is elementary matter
charged under the confined gauge group, is constructed by field theory
anomaly matching.  Once the massive ($m\ge\mu_c$) composite degrees of
freedom are integrated out, this generates a potential for the dilaton
and moduli.

The gaugino masses were given in~\cite{gnw}.  In the notation adopted here they
take the form\footnote{As in the above subsections we set $p_i=0$ in (\ref{vgs});
modifications that occur when $p_i\ne0$ are discussed in the following subsection.}
\beq M_a = {g^2_a(\mu)\over2}F^S + M\upp_a, \eeq
where $M\upp_a$ is given in (\ref{MlGSth}).  The A-terms, squared soft scalar masses 
and B-terms are given by (\ref{Atot}), (\ref{orbsm})--(\ref{orbsm1}) and (\ref{Btot})
respectively, with (see Appendix~A)
\bea
M = {1\over2}b_+^0u = - 3\mG, \quad F^S = - {1\over4}K^{-1}_{S\S}\(1+{g^2_s\over3}\bpl\)\u,
\quad K_S = - {1\over2}g^2_s, \label{ksBGW} \eea
where $g^2_s$ is defined in (\ref{gstr}) and $\bpl$ is the beta function
coefficient, Eq. (\ref{ba}), of the condensing gauge group $\G_+$.\footnote{If there
are several condensing gauge groups, the one with the largest value of $b_a^0$
dominates \susy breaking.}  The model of Ref.~\cite{bgw}
is explicitly modular invariant, so the moduli are stabilized at their self-dual points
with $\lvev F^\alpha\rvev = 0$, and supersymmetry breaking is dilaton dominated. 
Then~\cite{bgw}
\bea A\up_{ijk} = {g^2_s\over2}F^S, \quad M\up_i = {1\over3}|M| = |\mG|. \eea
Vanishing of the vacuum energy (2.4) now requires 
\beq    K_{S\S}|F^S|^2 = {1\over3}|M^2|, \quad K^{-1}_{S\S} = {(2\bpl)^2\over3
(1+{1\over3}g^2_s\bpl)^2}, \quad \left|{F^S\over M}\right| = {2\bpl\over3
(1+{1\over3}g^2_s\bpl)},\label{vac}\eeq
If $\bpl\ll 1$ the tree level A-terms and gaugino masses are suppressed 
relative to the the gravitino mass, whereas the scalar masses and
B-terms, $B\up_{ij}\approx-\mG$, are not.
Therefore one loop corrections can be neglected for the latter, but may be important for
the former.  It is clear from (\ref{MlGSth}) and (\ref{Atot}) that the
dominant one loop
corrections in this case are just the ``anomaly mediated'' terms found in~\cite{rs,hit}:
\bea M_a & \approx& M\up_a + g_a(\mu^2){b^0_a\over3}\bM = g_a(\mu^2)\mG\({\bpl\over
1+ {1\over3}g_s^2\bpl} - b^0_a\), \nonumber \\
A_{ijk} &\approx& A\up_{ijk} - {1\over3}M\(\gamma_i + \gamma_j + \gamma_k\) =
\mG\({g^2_s\bpl\over1+ {1\over3}g_s^2\bpl}  + \gamma_i + \gamma_j + \gamma_k\). \eea
This model was analyzed in detail in~\cite{gn2}. Over most of the allowed 
parameter space, $.1 \ge\bpl\gg b^0_a$, the tree contributions dominate.  However there
is a small region of parameter space with a sufficiently small value of $\bpl$ that
the gaugino masses and A-terms are similar to those in an ``anomaly mediated'' 
scenario~\cite{rs,hit,PoRa}.

\begin{figure}[t]
    \begin{center}
\centerline{
       \epsfig{file=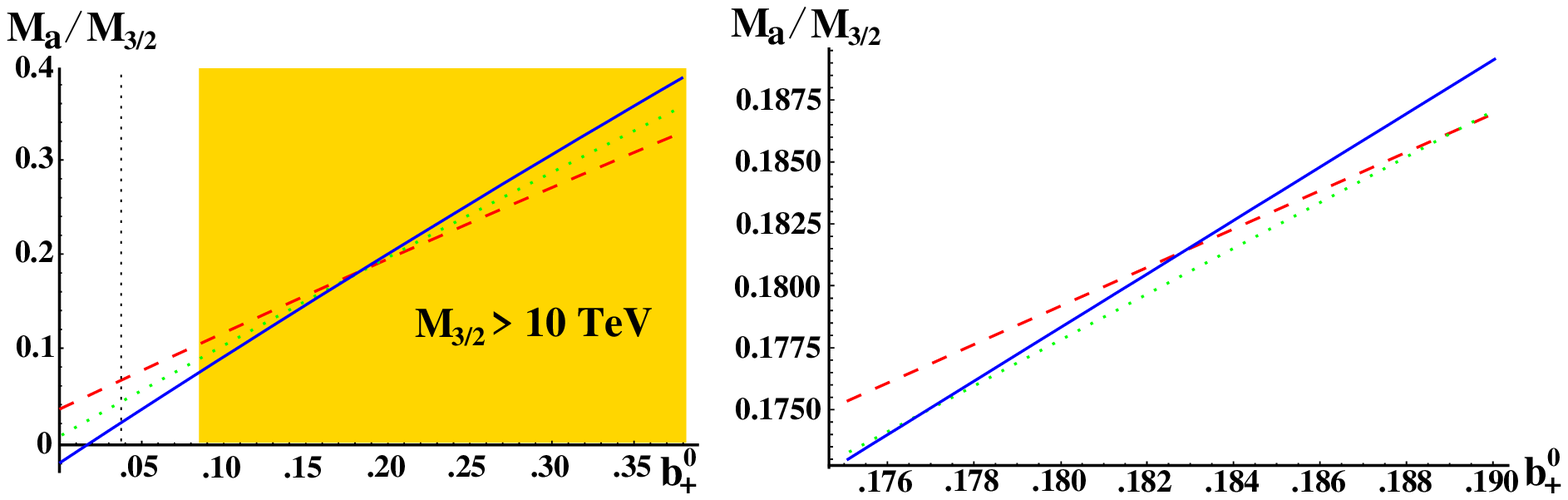,width=1.0\textwidth}}
          \caption{{\footnotesize {\bf Gaugino Masses in the BGW
                Model.} Gaugino masses $M_{1}$ (dashed), $M_{2}$
              (dotted) and $M_{3}$ (solid) are given at a scale
              $\Lambda_{\rm UV}= 1 \times 10^{14}$  GeV as a function
              of the condensing group beta function coefficient
              $b_{+}^{0}$. The  vertical dotted line in the left panel is the
              case of $E_{6}$ condensation in the hidden sector with 9
              {\bf 27}s of hidden sector matter studied
              in~\cite{gn2}. The right panel focuses on the region
              where the three masses are approximately unified.}}
        \label{fig:bgwgaug}
    \end{center}
\end{figure}

Using the expressions in Appendix~B, together with~(\ref{ksBGW})
and~(\ref{vac}) the pattern of soft supersymmetry breaking terms can
be obtained as a function of the condensing group beta function
coefficient $b_{+}^{0}$ and the modular weights of the
fields with $\lang {\rm Re}\;t \rang = 1$ or $\lang {\rm Re}\; t \rang =
e^{i\pi/6}$ and $\sin\theta = 1$. The condensation scale in these
models is typically of the
order of $1 \times 10^{14}$ GeV and we take this to be the boundary
condition scale $\Lambda_{\rm UV}$ in what follows. In
Figure~\ref{fig:bgwgaug} the
gaugino masses are displayed as a function $b_{+}^{0}$ as a fraction
of the gravitino mass. In~\cite{gn2}
it was shown that for weak coupling at the string scale ($g_{s}^{2}
\approx 1/2$) a reasonable scale of supersymmetry breaking ({\em i.e.}
gravitino masses less than 10 TeV) generally requires $b_{+}^{0} \leq
0.085$. The region with gravitino mass larger than $10$ TeV is shaded
in Figure~\ref{fig:bgwgaug}. Also
indicated in Figure~\ref{fig:bgwgaug} is a benchmark scenario
consisting of an $E_{6}$ gaugino condensate in the hidden sector
together with 9 {\bf 27}s of matter and having a beta function
coefficient of $b_{+}^{0} = 0.038$. 

The spectrum of gaugino masses will typically be similar to that of
the ``anomaly-mediated'' cases with $M_{1} \geq M_{2}$ and a lightest
neutralino with substantial wino-like content provided $b_{+}^{0} \leq
0.19$. The location of
the approximate unification of gaugino masses near this value of
$b_{+}^{0}$ is expanded in the right panel of Figure~\ref{fig:bgwgaug}.

\begin{figure}[t]
    \begin{center}
\centerline{
       \epsfig{file=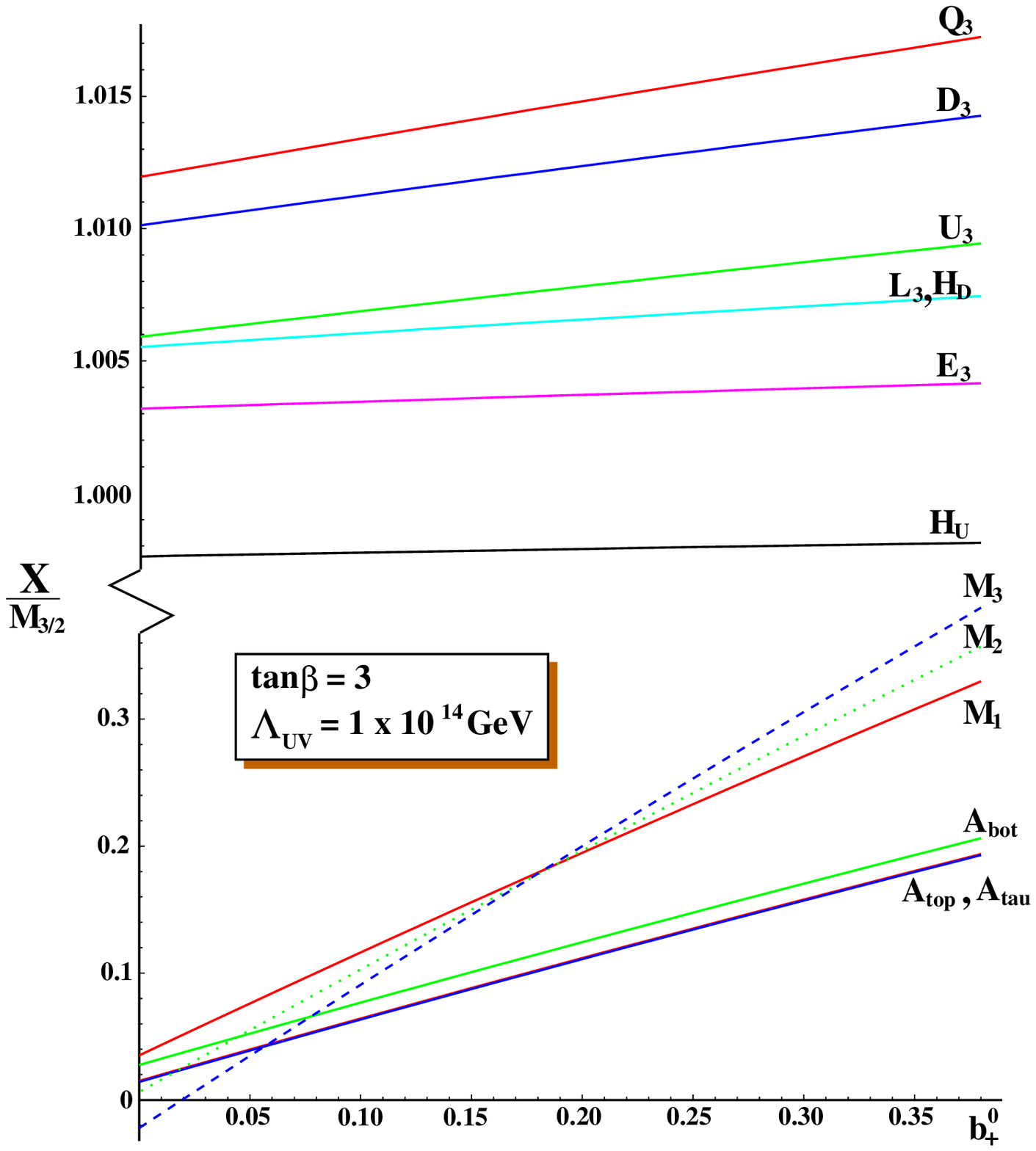,width=0.9\textwidth}}
          \caption{{\footnotesize {\bf Spectrum of Soft Supersymmetry
                Breaking Terms in BGW Model.} All values are given
              relative to the gravitino mass $m_{3/2}$ at a scale
              $\Lambda_{\rm UV}= 1 \times 10^{14}$ GeV as a function
              of the condensing group beta function coefficient $b_{+}^{0}$.}}
        \label{fig:bgwspectrum}
    \end{center}
\end{figure}

In Figure~\ref{fig:bgwspectrum} we plot the relative sizes of all
third generation scalar masses and A-terms, Higgs masses and gaugino
masses as a fraction of the gravitino mass for $\tan\beta =3$,
assuming $n_{i} = -1$ for all fields. As was
the case in
Sections~\ref{sec:anomalycase}-\ref{sec:O1case}, the gauginos are
typically an order of magnitude smaller than scalars (note the change
in vertical scale in Figure~\ref{fig:bgwspectrum}). Despite this
hierarchy, this model was shown in~\cite{gn2} to give rise to
acceptable low energy phenomenology provided $\tan\beta$ was in the
low to moderate range.  Figure~\ref{fig:bgwspectrum} displays an
important feature of the always-present one loop contributions arising
from the conformal anomaly: when tree level scalar masses are present
and universal the non-universality arising from the anomaly pieces is
negligible (here averaging less than a 1\% correction). However, the
corrections to the gaugino masses may significantly alter the gaugino
spectrum {\em provided the tree level contributions are absent or
  suppressed}, as in the BGW model considered here. Neglecting these
one loop anomaly-induced contributions to soft terms is an
approximation whose validity needs to be assessed on a model-by-model basis.

\subsection{Matter couplings to the Green-Schwarz term}

So far we have assumed the Green-Schwarz function $V_{GS}$ depends only on the
moduli, that is, we set $p_i = 0$ in (\ref{vgs}).  Only the moduli couplings in this
term are known from string loop calculations~\cite{dkl,ant} and they are proportional
to the K\"ahler potential for the moduli.  It is possible that the GS function is 
proportional to the full K\"alher potential, in which case $p_i = p = -\dgs/24\pi^2$,
or that it is proportional to the untwisted K\"ahler potential, {\it i.e.} to the logarithm 
of the determinant of the metric in the six dimensional compact space.  In this
last case we would have $p_i = p$ for untwisted matter and $p_i = 0$ for twisted matter.
The presence of these terms modifies the soft parameters if $F^S\ne 0$.

One effect of $p_i\ne0$ is a modification~\cite{adam} of the
``effective'' matter K\"ahler potential (\ref{kappa}): \bea \kappa_i
&\to& \(1 + {1\over2}g_sp_i\)\kappa_i.\label{keff} \eea The potential
can still be written in the form given in (A.8) of the appendix with
the replacement $K_{N\bN} \to \hat{K}_{N\bN} = K_{N\bN} +
{1\over2}g_s(V_{GS})_{N\bN}$.  However the effective metric is not
K\"ahler in this formulation.  In addition $F^N$ does not take the
usual form (\ref{F}): $F^N = - e^{-K/2}W^{-1}\hat{K}^{N\bN}\pp_{\bN}
\(e^KW\W\)$, which reduces to (\ref{F}) when $W$ is holomorphic.  This
is not the case in the linear multiplet formulation for the dilaton
that we are using here because of the way the GS term enters in the
dilaton potential, as described in Appendix A.  For these reasons
Eqs. (\ref{aijk}), (\ref{smass}) and (\ref{bij}) do not generally
apply if $F^S\ne 0$; the $p_i$-terms in these parameters depend on the
specifics of the model for generating a potential for the dilaton.
The PV metrics (\ref{kapPhi}) are similarly modified: \beq
\kappa^\Phi_i \to \(1 + {1\over2}g_sp_i\)\kappa^\Phi_i, \quad
\hka^\Phi_i \to \(1 + {1\over2}g_sp_i\)^{-1}\hka^\Phi_i, \eeq as are
the soft parameters in the PV potential.  These give additional
one loop contributions, which can be important for gaugino masses
which have no tree level contribution from $p_i\ne0$.

Here we give the results only for the explicit dilaton dominated \susy
breaking model of the 
previous subsection: 
\bea \Delta A_{ijk} &\approx& \Delta A_{ijk}\up = 
-{p_i\(3 + g^2_s\bpl\)\over2\bpl\(1 + {1\over2}g_sp_i\)}\mG + (i\to j) + (j\to k), \nonumber \\
 \Delta B_{ij} &\approx&  \Delta B_{ij}\up
- {p_i\(3 + g^2_s\bpl\)\over2\bpl\(1 + {1\over2}g_sp_i\)}\mG
+ (i\to j), \nonumber \\ \Delta M_a &\approx& \Delta M_a\upp = {g^2(\mu)\over8\pi^2}\sum_i
{C^i_ap_i\(3 + g^2_s\bpl\)\over2\bpl\(1 + {1\over2}g^2_sp_i\)}\mG.\label{pterms}\eea
Note that in this special case the above results can in fact
be obtained from the general formulae (\ref{MlGSth}), (\ref{aijk}) and (\ref{bij}):
\bea \Delta A\up_{ijk} &=& - F^SK_{S\S}{p_i\over1 + {1\over2}g_sp_i} +
(i\to j) + (j\to k), 
\nonumber \\
 \Delta B\up_{ij} &=& - F^SK_{S\S}{p_i\over1 + {1\over2}g_sp_i} + (i\to j), \nonumber \\
 \Delta M\upp_a &=&
{g^2(\mu)\over8\pi^2}F^SK_{S\S}\sum_i{p_iC^i_a\over1 
+ {1\over2}g^2_sp_i},
\eea
since it follows from (\ref{keff}) that 
\bea
F^n\pp_n\ln\kappa_i &\to& F^n\pp_n\ln\kappa_i - F^SK_{S\S}{p_i\over1 + {1\over2}g_sp_i},\eea
where we used the relation
\beq {\pp g_s\over\pp s} = 2{\pp\ell\over\pp x} = - K_{S\S},\eeq
given in Appendix A. However (\ref{smass}) does not apply even in this case;
the tree level scalar masses in this model have been given in~\cite{bgw}:
\beq |M\up_i| = {1\over\bpl}\left|{3p_i - 2\bpl\over2 + p_ig^2_s}\mG\right|.\eeq
The results (\ref{pterms}) then follow from (\ref{vac}).
We see a considerable enhancement of all these parameters if $p_i=p>>\bpl$.  Under the
assumption that $-\dgs$ takes its maximum value of 90, the only viable
scenario with some $p_i = p$ found in~\cite{gn2} is for $p_{H_{u,d}}= 0$ and
$p_i = p$ for all three generations of squarks and sleptons.

\vskip .5cm
\section{Conclusion}
To conclude, let us first stress that even though we have been
studying specific classes of superstring models, the types of spectra
that we obtained and discussed appear to be quite generic. For example,
scenarios from models with extra dimensions tend to give spectra which
can be related to one or another type considered here, whether it is the
model of Randall and Sundrum \cite{rs}, or models of gaugino mediation 
\cite{gauginomed}.

In particular, soft terms that are proportional to beta function
coefficients and anomalous dimensions can be realized in a variety of
ways in string-derived supergravity. The case that is generally
referred to as ``anomaly mediation'' is just one limiting value in a
continuum of such models. The importance of these anomaly-induced
terms depends on the absence or suppression of tree level
contributions to the soft supersymmetry breaking parameters and on the
assumptions made regarding the underlying theory when regulating the
effective supergravity theory. 

Once supersymmetry is discovered, the central issue will be to
unravel the mechanism of supersymmetry breaking. The search strategy
will be of the most value if it is based on large classes of different
models, not just on a single ``minimal'' model. The models studied above
tend to show that a possible strategy could be based on three steps:

(i) identifying gaugino masses (the least model dependent aspect of
these theories) and the nature of the LSP,

(ii) identifying where (approximately) the bulk of the scalar masses lie
and whether there is an order of magnitude between gaugino and scalar
masses, 

(iii) then using the detail of the scalar masses, in particular the
mixing in the stop sector and the degree of
non-universality, to disentangle the possible scenarios.

Observation of non-universal supersymmetric
parameters obeying the relations described in
Sections~\ref{sec:anomalycase}-\ref{BGWcase} will likely shed
light on the scale of supersymmetry breaking, the nature of the fields
responsible for this breaking and the origin of the $\mu$-term, if not
the properties of the underlying superstring theory itself.

\vskip .5cm
\noindent {\bf Acknowledgements}
\vskip .5cm We thank Joel Giedt for discussions. P.B. thanks the Theory
Group of LBNL for its generous hospitality and the participants of the
GDR Supersymmetry working group on non-universalities, especially
Laurent Duflot and Jean-Fran\c{c}ois Grivaz, for discussions.
B.N. would like to thank the Laboratoire de Physique Th\'eorique at
the Universit\'e Paris-Sud where part of this work was completed. This work was
supported in part by the Director, Office of Science, Office of Basic
Energy Services, of the U.S. Department of Energy under Contract
DE-AC03-76SF00098 and in part by the National Science Foundation under
grants PHY-95-14797 and INT-9910077.

\vskip .5cm
\appendix
\noindent{\large \bf Appendix}
\def\ksubsection{\Alph{subsection}}
\def\theequation{\ksubsection.\arabic{equation}} 

%       reset section commands
     
\catcode`\@=11

\def\thesubsection{\Alph{subsection}.}
\def\thesubsubsection{\arabic{subsubsection}.}

\subsection{The linear multiplet formalism for the dilaton}
\setcounter{equation}{0}
\label{AppendixLin}

In this paper we have presented the soft supersymmetry breaking parameters 
in terms of the various auxiliary fields of supergravity. In order
to adhere as closely as possible to the notation of~\cite{bim}, we used
expressions of the form obtained in the standard chiral formulation of
supergravity. In the context of string theory, the dilaton $\ell$
appears as the scalar component of a linear multiplet $L$.  The chiral
multiplet formulation can be recovered by a duality transformation, at
least at the classical level.  However the linear multiplet formulation
provides a simpler implementation of the Green-Schwarz anomaly
cancellation mechanism and a better framework for constructing an
effective Lagrangian for gaugino condensation.  The effective theory
of~\cite{bgw} made explicit use of the linear multiplet formalism.  In
this appendix we show the correspondence between various terms in the
component Lagrangian of the linear formalism and of the expressions
given in the text.  We also show how explicit cancellations among the
light loop (anomaly) contribution, the GS term and the string threshold
corrections result in the final expression (\ref{MlGSth}) for the
one loop gaugino mass. These cancellations are most readily displayed in
the linear multiplet formalism.  Finally, we will display corrections to
the soft parameters in the scalar potential that are present if
the dilaton and moduli sectors both contribute substantially to supersymmetry 
breaking.

In the presence of a (nonperturbatively induced) potential for the dilaton, 
the tree level scalar Lagrangian takes the form (dropping gauge charged matter)
\beq \L_{\rm scalar} = - \sum_\alpha {\pp_m\ta\pp^m\bta\over(\ta + \bta)^2} -
{k'(\ell)\over4\ell}\pp_m\ell\pp^m\ell - {\ell\over
k'(\ell)}\pp_m a\pp^m a - V, \label{ls} \eeq
where the axion $a$ is related to the two-form $b_{mn}$ of the linear
multiplet by a duality transformation:
\beq {1\over2}\epsilon^{mnpq}\pp_nb_{pq} = - {2\ell\over k'(\ell)}\pp^ma.\eeq
The potential $V$ can be written in the form
\bea 
V &=& \sum_\alpha{1\over(\ta + \ta)^2} F^\alpha\bF^\alpha 
+ {\ell\over k'(\ell)}F^2 - {1\over3}M\bM, \nonumber \\ 
F &=& {k'(\ell)\over 4\ell}f(\ell,\ta,z^i),\label{vdil}
\eea
where $f(\ell,\ta,z^i)$ is a complex but nonholomorphic function of the
scalar fields.  For example in the model of~\cite{bgw},
\beq f(\ell,\ta,z^i) = - \sum_a(1+ \ell b_a)\bar{u}_a \approx -
(1+ \ell b_+)\bar{u}_+,\label{bgw}\eeq
where $\bar{u}_a(\ell,\ta,z^i)$ is the value of the gaugino condensate
for a hidden gauge group $\G_a$ with beta function coefficient $b_a =
\(C_a - {1\over3}\sum_iC_a^i\)/8\pi^2 = 2b_a^0/3$; the function (\ref{bgw}) is dominated by the
condensate $\bar{u}_+$ with the largest beta function coefficient $b_+$.

To cast this result in a form resembling the
standard chiral formulation we introduce the variable $x(\ell) = 2
g^{-2}(M_s)$, which is twice the inverse squared gauge couplng (\ref{gstr}). It
is related to the dilaton K\"ahler potential $k$ by the differential
equation~\cite{physrep}
\beq k'(\ell) = -\ell x'(\ell), \quad \pp\ell = - {\ell\over
k'(\ell)}\pp x, \eeq
giving
\bea &&{\pp k(x)\over\pp x} = k'(\ell){\pp\ell\over\pp x} = - \ell, \quad
{\pp^2 k(x)\over\pp x^2} = -{\pp\ell\over\pp x} = {\ell\over
k'(\ell)}, \nonumber \\&&
 {k'(\ell)\over4\ell}\pp_m\ell\pp^m\ell = -{\ell\over4k'(\ell)}\pp_mx\pp^mx
= {1\over4}{\pp^2 k(x)\over\pp x^2}\pp_mx\pp^mx.  \eea
Then setting 
\beq x = s + \s, \quad a = {\im s}, \eeq
(\ref{ls}) and (\ref{vdil}) take the standard form (including gauge-charged chiral matter)
\bea \L_{\rm scalar} &=& - \sum_NK_{N\bN}\(\pp_m z^N\pp^m\z^{\bN} 
+ F^N\bF^{\bN}\) + {1\over3}M\bM, \nonumber \\
K &=& k\(s + \s\) + K(\ta) + \sum_i\kappa_i|z^i|^2,\label{chiral} \eea
provided we identify $F = F^S$ and $K_{S\S} = \ell/k'(\ell)$.
When the fermion part of the
Lagrangian is included, one obtains for the gaugino masses 
\beq M\up_a = {g^2_a\over2}F, \eeq
in agreement with (\ref{M0}) with $f_a = s$ and $F = F^S$.
 
The replacements (A.7) amount to a duality transformation to the chiral
formulation for the dilaton.  When the GS term is included, after 
a two-form/scalar duality transformation, Eqs.  (\ref{ls})--(\ref{chiral})
are modified by the replacements 
\bea \pp_ma&\to&\pp_ma +
{b\over2}\sum_\alpha{\pp_m\im\ta\over\re\ta}, \quad
\(\ta + \bta\)^{-2}\to \(1 + b\ell\)\(\ta +
\bta\)^{-2}, \nonumber \\ b &=& -\dgs/24\pi^2. \label{modify} \eea
  We may make a full superfield duality 
transformation by the additional replacements 
\beq x = s + \s = \ts + \bts + b\sum_\alpha\ln(\ta + \bta), \quad 
k(s+\s)\to  k\[\ts + \bts - b K(\ta)\]\label{modify2},\eeq
where $\ts$ is the complex scalar component ($\im\ts = a$) of the dilaton chiral
superfield.
This introduces mixing of the moduli [and of  matter fields if $p_i\ne0$ in (\ref{vgs})]
with the dilaton in the K\"ahler metric~\cite{bim}.
Working in the linear multiplet formalism for the dilaton, there is no mixing
of the dilaton with chiral fields;\footnote{See
for example the discussion of Eq. (4.20) in~\cite{tom}.} in this case (\ref{ls})
and (\ref{vdil}) are modified only by (\ref{modify}).
With this modification (\ref{vdil}) is completely general; it includes the effects of the
GS term on the potential for $\ell$ and $t$ in the presence of 
a source of \susy~breaking such as gaugino condensation.  In
fact the GS term coupling to the confined hidden gauge sector, as in
the model of Section~\ref{BGWcase}, must be included to make the
effective \susy breaking ``tree'' Lagrangian perturbatively modular invariant.
 
However it is inconsistent to include the GS term coupling to the
unconfined (observable) gauge sector without the corrections from the
observable sector loops. 
Here we illustrate the modular anomaly cancellation
among the contributions to the gaugino masses.
In orbifold models the light loop contribution (\ref{Man}) takes the form
\bea
M^{(1)}_a|_{\rm an} &=& {g^2_a(\mu)\over 2}\lbr{2b^0_a\over3}\bM
+ {\ell\over8\pi^2}\(C_a - \sum_iC^i_a\)F\right.
\nonumber \\ & & \qquad \left. + \sum_\alpha F^\alpha{2\over3(\ta + \bta)}\[
b^0_a - {1 \over8\pi^2} \sum_i C^i_a(1+ 3n_i)\]\rbr, \label{Manorb}\eea
The contribution of the GS term (\ref{GS}) is
\beq M^{(1)}_a|_{\rm GS} = {g^2_a(\mu)\over 2}\sum_\alpha F^\alpha
{2\over3(\ta + \bta)}{\dgs\over16\pi^2}.\label{MGS}\eeq
and the string threshold corrections (\ref{vgs}) give a contribution
\beq M^{(1)}_a|_{\rm th} = {g^2_a(\mu)\over 2}\sum_\alpha F^\alpha
{4\over3}\zeta(\ta)\[{\dgs\over16\pi^2} b^0_a - 
{1 \over8\pi^2} \sum_i C^i_a(1+ 3n_i)\].\label{Mth}\eeq
These combine to give the total contribution
(\ref{MlGSth}) with the substitutions 
$$F\to F^S,\quad\ell\to g^2_s/2= - K_S,$$
with the moduli $\ta$ appearing only through the modular invariant
expressions $$F^\alpha\[(\ta + \bta)^{-1} + 2\zeta(\ta)\].$$

In the linear multiplet formulation for the dilaton, the 
tree level scalar potential takes the form 
\bea V_{\rm tree} &=& \sum_N\hK_{N\bN}F^N\bF^{\bN} - {1\over3}M\bM, 
\nonumber \\ M &=& - 3e^{K/2}w, \quad F^N = - w^{-1}e^{-K/2}\hK^{N\bN}\pp_{\bN}\(e^Kw\w\),
\label{linpot}\eea
where the effective metric
$\hK_{N\bN}$ is defined in (\ref{khat}), and $\hK_{S\S} = K_{S\S} = \ell/k'(\ell)$.
(\ref{linpot}) reduces to the standard form if $w$ is holomorphic.
If a duality transformation to the chiral formulation for the dilaton is always
possible~\cite{BDQQ} in the effective theory below the \susy breaking scale, we must
have 
\bea w &=& w(\ts,\ta,z^i), \quad s = {x\over2} + ia,\quad \ts = s + {1\over2}\vgs.\eea
For example,
in the BGW model we have\footnote{The full potential for the BGW model is given in (15) 
of~\cite{glm}.  The full expression for the field dependence of the condensate $u$ with
$z^i=0$ is given in the second reference of~\cite{bgw}, and reduces to (\ref{cond}) with
the identification of the axion as $a = - b_+\omega$ in the notation of that paper.}
\beq w = W(\ta,z^i) + v(\ts,\ta), \quad v = - e^{-K/2}{b_+\over4}u,\quad
u = ce^{K/2}e^{-\ts/b_+}\prod_\alpha\eta(\ta)^{2(b-b_+)/b_+},\label{cond}\eeq
where $c$ is a constant.  In this case we have 
\bea \pp_{\bN}w &=& {1\over2}w_S\pp_{\bN}\vgs, \quad w_{\S} = 0, \quad
F^S = -e^{K/2}\hK^{S\S}\[\w_{\S} + K_{\S}\w\] \nonumber \\
F^N &=& -e^{K/2}\hK^{N\bN}\[\w_{\bN} + K_{\bN}\w + {1\over2}(\pp_{\bN}\vgs)\pp_s\ln w\].
\eea
Inserting these expressions in the potential (\ref{linpot}) we obtain the 
following expressions for the soft \susy-breaking terms at tree level:
\bea A^{\rm tree}_{ijk} &=& A\up_{ijk} - {b\over2(\ta + \bta)}
\(F^\alpha\pp_s\ln\w + {\rm h.c.}\), \nonumber \\
 \[B_{ij}^{\rm tree}\]_{\rm superpotential} &=& \[B\up_{ij}\]_{\rm superpotential} 
- {b\over2(\ta + \bta)}
\(F^\alpha\pp_s\ln\w + {\rm h.c.}\), \nonumber \\
 \[B_{ij}^{\rm tree}\]_{\rm K\ddot{a}hler\; potential} &=& 
\[B\up_{ij}\]_{\rm K\ddot{a}hler\; potential} 
- {b\over2(\ta + \bta)}F^\alpha\pp_s\ln\w,\label{vijk}\eea
where the expressions with index 0 are the tree level expressions given in the
text with $W(Z^N)\to w(Z^N,\vgs)$ and
\beq F^\alpha = -e^{K/2}\hK^{\ta\bta}\[\w_{\bta} + K_{\bta}\w -
{b\over2(\ta + \bta)}\pp_s\ln w\]\label{newft}.\eeq
The scalar masses depend on the curvature of the effective
scalar metric $\hK_{N\bN}$.  If $p_i\ne0$ they are complicated expressions
in the general case; their values for the BGW model are given in
Section~3.5. If $p_i=0$, they reduce to the result given in Section~2.4, with
the substitutions $W\to w$ and (\ref{newft}). 

If $p_i = 0$ the expressions in Section 2 receive no corrections if
\susy breaking is dilaton mediated, $F^\alpha = 0$. If there is no
dilaton ``superpotential'', $w_s = 0$, the only correction is the
rescaling $F^\alpha \to (1 + b\ell)F^\alpha$.  If a dilaton
``superpotential'' $v$ is generated by a single dominant gaugino
condensate (and  the associated matter condensates), the dilaton
dependence of $v$ in (\ref{cond}) follows quite generally from anomaly
matching, giving \beq \pp_s\ln w = v/b_+w.\eeq Since $b_+\le b$, the
corrections in (\ref{vijk}) can be significant if $|v/w|,\cos\theta$
and $1/\ta$ are all order one.  The moduli dependence of $v$ in
(\ref{cond}) follows from perturbative modular
invariance.\footnote{Modular invariance could be broken in $v$ if
corrections to $k(\ell)$ from string nonperturbative effects are
moduli dependent~\cite{eva}.  We have ignored this possibility
throughout.}  To the extent that modular invariant condensation
dominates \susy~breaking, one gets essentially the BGW model with
negligible contributions from $F^\alpha$. On the other hand if $\lvev
W\rvev$ is dominant, the corrections found in (\ref{vijk}) again
become negligible.  They are
significant only if there are two comparable sources of
\susy~breaking.  Even in this case they are unimportant in the large
$T$ limit if $\pp_s\ln w$ is not too large.  Note that the correction
to the A-term does not vanish at the self-dual points for the moduli,
so in this case we would not get an ``anomaly mediated'' scenario at
these points when $F^S=0$.  In the chiral formulation~\cite{bim},
there is mixing between dilaton and moduli F-terms.  In that language,
the corrections to the results of Section~2, aside from the rescaling
of $F^\alpha$, arise from terms proportional to
$F^S\bF^{\alpha}K_{S\T^\alpha} +$ h.c.  in the potential.

\subsection{Soft \susy~breaking terms in orbifold models}
\setcounter{equation}{0}
\label{AppendixSoft}
In this appendix we collect the complete expressions (tree plus
one loop correction) for the soft
supersymmetry breaking terms in orbifold models defined
by~(\ref{f=S}), (\ref{Kahler}) and~(\ref{kap}) with supersymmetry
breaking {\em vevs} parameterized by~(\ref{sin}) and~(\ref{alphacos}).
We neglect corrections proportional to $\dgs/48\pi^2$ in the scalar
potential that were discussed in Appendix A.

The gaugino mass is determined from~(\ref{orbma}) and~(\ref{MlGSth}):
\begin{eqnarray}
M_{a}^{tot}&=&\frac{g_{a}^{2}(\mu)}{2\sqrt{3}}\overline{M}\lbr
\frac{2}{3}\cos\theta\sum_{\alpha} (t^{\alpha}+
\overline{t}^{\alpha}) G^{\alpha}_{2}
\Theta_{\alpha}\[ \frac{\delta_{\rm GS}}{16\pi^{2}}
+ b_{a}^{0}-\frac{1}{8\pi^{2}}\sum_{i}C_{a}^{i}(1+3n^{\alpha}_{i})\]
e^{-i \gamma_{T}} 
 \right. \nonumber \\
& & \left. +\frac{2 b_{a}^{0}}{\sqrt{3}} +
\frac{\sin\theta}{k_{s\overline{s}}^{1/2}}\[ 1+\frac{g_{s}^{2}}{16\pi^2}\(C_a
-\sum_{i}C_{a}^{i}\)\]  e^{-i \gamma_{S}} \rbr.
\label{MaFull}
\end{eqnarray}
The trilinear A-terms are obtained from~(\ref{Atot}). The expression
is simplified by utilizing~(\ref{diag}) to
obtain the identities
\begin{eqnarray}
F^{S}\partial_{s}\gamma_{i}^{a}=-\gamma_{i}^{a} M_{a}^{(0)};&
\partial_{s}\gamma_{i}^{lm}=k_{s}\gamma_{i}^{lm},
\label{gammagauge}
\end{eqnarray}
where the last relation is true if $\kappa_{i} \not=
\kappa_{i}(s)$. This yields A-terms of the form:
%*** This allows a bit of consolidation between the second line and the
%last line of~(\ref{Atot})) Note that the
%A-term pieces coming from $F^{S}$ cancel, but the gaugino
%mass terms do not, due to that pesky relative factor of two. If it
%weren't for the factor of 2 the $\ln(\tmu^2_{ia})$ term in~(\ref{AtermO2A})
%would vanish. A similar thing happens with the B-terms. Also, I
%switched an $M$ for an $\overline{M}$ in the anomaly-induced
%piece. Was it supposed to be an $\overline{M}$ all along?***
\begin{footnotesize}
\begin{eqnarray}
A^{tot}_{ijk} &=& \frac{\overline{M}}{\sqrt{3}} \lbr
-\frac{\gamma_{i}}{\sqrt{3}} +\cos\theta \[\sum_{\alpha}
 (t^{\alpha}+\overline{t}^{\alpha}) G^{\alpha}_{2} \Theta_{\alpha} \(
\frac{1}{3} (n^{\alpha}_{i}+ n^{\alpha}_{j} + n^{\alpha}_{k} +1)
 - \sum_{lm} \gamma_{i}^{lm}
\( p_{lm}^{\alpha} -  (n^{\alpha}_{i}+ n^{\alpha}_{l} + n^{\alpha}_{m}
+1) \ln(\tmu^2_{lm}) \) \right. \right. \right. \nonumber \\
 & & \left. \left. \left. -\sum_{a} \gamma_{i}^{a} p_{ia}^{\alpha}\)
       -\sum_{\beta} 
     \ln\[(t^{\beta}+\overline{t}^{\beta}) |\eta(t^{\beta})|^4\]
     \sum_{lm} \gamma_{i}^{lm} p_{lm}^{\beta} \sum_{\alpha}
     (t^{\alpha} + \overline{t}^{\alpha}) G^{\alpha}_{2}
 \Theta_{\alpha} (n^{\alpha}_{i}+
n^{\alpha}_{l} + n^{\alpha}_{m} +1) \] e^{-i \gamma_{\alpha}} \right. \nonumber \\
 & & \left. + \frac{\sin\theta}{k_{s\overline{s}}^{1/2}}
   \[-\frac{k_{s}}{3} +\sum_{a}\gamma_{i}^{a} \( \partial_{s}
   \ln(\tmu^2_{ia}) + \frac{g_{a}^{2}}{2}\ln(\tmu^2_{ia}) \) +\sum_{lm}
   \gamma_{i}^{lm} \partial_{s} \ln(\tmu^2_{lm})  \right. \right. \nonumber
\\
& & \left. \left. -\sum_{\alpha}
     \ln\[(t^{\alpha}+\overline{t}^{\alpha}) |\eta(t^{\alpha})|^4\] \(
     \sum_{a} g_{a}^{2} \gamma_{i}^{a} p_{ia}^{\alpha} - k_{s} \sum_{lm}
     \gamma_{i}^{lm} p_{lm}^{\alpha} \) \]  e^{-i \gamma_{S}} \rbr
     +\rm{cyclic}(ijk).  
\label{Atermfull}
\end{eqnarray}
\end{footnotesize}
The bilinear B-terms have a similar form
\begin{footnotesize}
\begin{eqnarray}
B_{ij}^{tot}&=& \frac{\overline{M}}{3}\(\frac{1}{2} - \gamma_{i}\) +
\frac{\overline{M}}{\sqrt{3}} \lbr \cos\theta \[
\frac{1}{2}\sum_{\alpha} \Theta_{\alpha} \( (n_{i}^{\alpha} +
n_{j}^{\alpha} +1)-\partial_{t^{\alpha}} \ln \mu_{ij} \)
-\sum_{\alpha} (t^{\alpha} +
\overline{t}^{\alpha}) G_{2}^{\alpha} \Theta_{\alpha}  \( \sum_{a}
\gamma_{i}^{a} 
 p_{ia}^{\alpha} + \sum_{lm} \gamma_{i}^{lm} p_{lm}^{\alpha} \)
\right. \right. \nonumber \\
 & & \left. \left. -\sum_{\beta}
     \ln\[(t^{\beta}+\overline{t}^{\beta}) |\eta(t^{\beta})|^4\]
     \sum_{lm} \gamma_{i}^{lm} p_{lm}^{\beta} \sum_{\alpha}
     (t^{\alpha}+\overline{t}^{\alpha}) G_{2}^{\alpha} \Theta_{\alpha}
     (n^{\alpha}_{i}+ n^{\alpha}_{l} + n^{\alpha}_{m} +1)
   \right. \right. \nonumber \\
& & \left. \left. +\sum_{lm} \gamma_{i}^{lm} \sum_{\alpha}
     (t^{\alpha}+\overline{t}^{\alpha}) G_{2}^{\alpha} \Theta_{\alpha}
     (n^{\alpha}_{i}+ n^{\alpha}_{l} + n^{\alpha}_{m} +1)
     \ln(\tmu^2_{lm})\] e^{-i\gamma_{\alpha}}
       +\frac{\sin\theta}{k_{s\overline{s}}^{1/2}}  \[ \sum_{a} \gamma_{i}^{a} \(\frac{g_{a}^{s}}{2}
    \ln(\tmu^2_{ia}) + \partial_{s} \ln(\tmu^2_{ia}) \)  \right. \right. \nonumber \\
& & \left. \left.  +\sum_{lm} \gamma_{i}^{lm}
    \partial_{s} \ln(\tmu^2_{lm}) -\sum_{\alpha}
    \ln\[(t^{\alpha}+\overline{t}^{\alpha}) |\eta(t^{\alpha})|^4\] \(
    \sum_{a} \gamma_{i}^{a} g_{a}^{2} p_{ia}^{\alpha} - \sum_{lm} \gamma_{i}^{lm}
    k_{s} p_{lm}^{\alpha} \) -(k_{s}
  +\partial_{s} \ln \mu_{ij} ) \] e^{-i\gamma_{s}} \rbr 
\nonumber \\ & & + (i \leftrightarrow j)
\label{Btermfull}
\end{eqnarray}
\end{footnotesize}

The scalar masses arise from~(\ref{orbsm}) and~(\ref{orbsm1}). Some
degree of consolidation can be obtained by employing the
relation~(\ref{gammagauge}) as well as
\begin{equation}
|F^{S}|^{2}\partial_{s}\partial_{\overline{s}}\ln g_{a}^{2} =
|M_{a}^{(0)}|^{2}, 
\end{equation}
to allow the following identifications:
\begin{footnotesize}
\begin{eqnarray}
\overline{F}^{\overline{S}}\partial_{\overline{s}} \sum_{a}
\gamma_{i}^{a} M_{a}^{(0)} \ln(\tmu^2_{ia}) &=&
\sum_{a} \gamma_{i}^{a} \lbr \overline{F}^{\overline{S}} M_{a}^{(0)}
\partial_{\overline{s}} 
\ln(\tmu^2_{ia}) -2(M_{a}^{(0)})^{2}\ln(\tmu^2_{ia})\rbr \nonumber \\
\overline{F}^{\overline{S}}\partial_{\overline{s}} \sum_{lm}
\gamma_{i}^{lm} A_{ilm}^{(0)} \ln(\tmu^2_{lm}) &=& \sum_{lm}
\gamma_{i}^{lm} \lbr A_{ilm}^{(0)}
\overline{F}^{\overline{S}} \partial_{\overline{s}} \ln(\tmu^2_{lm}) +
 k_{\overline{s}} \overline{F}^{\overline{S}} A_{ilm}^{(0)}
  \ln(\tmu^2_{lm}) -k_{s\overline{s}} |F^{S}|^{2} \ln(\tmu^2_{lm})
  \rbr \nonumber \\
F^{S}\overline{F}^{\overline{S}} \partial_{s}\partial_{\overline{s}}
\(\sum_{a} \gamma_{i}^{a} \ln(\tmu^2_{ia})\) &=&\sum_{a} \gamma_{i}^{a} \lbr
2(M_{a}^{(0)})^{2}\ln(\tmu^2_{ia}) +|F^{S}|^{2}
\partial_{s}\partial_{\overline{s}} \ln (\tmu^2_{ia})
- \[M_{a}^{(0)} \overline{F}^{\overline{S}} \partial_{\overline{s}}
\ln(\tmu^2_{ia}) + {\rm h.c.}\] \rbr \nonumber \\
F^{S}\overline{F}^{\overline{S}} \partial_{s}\partial_{\overline{s}}
\(\sum_{lm} \gamma_{i}^{lm} \ln(\tmu^2_{lm})\) &=& |F^{S}|^{2} \sum_{lm} 
\gamma_{i}^{lm} \lbr \partial_{s}\partial_{\overline{s}} \ln(\tmu^2_{lm})
+\(k_{s\overline{s}} +k_{s}k_{\overline{s}}\) \ln(\tmu^2_{lm}) +
\[k_{s}\partial_{\overline{s}} \ln (\tmu^2_{lm}) + {\rm h.c.}\] \rbr
\nonumber .
\label{sderivs}
\end{eqnarray}
\end{footnotesize}
With these, the complete expression for the tree level plus one loop
scalar masses is given by
\begin{footnotesize}
\begin{eqnarray}
(M_{i}^{tot})^{2}&=&|M|^{2}\lbr\frac{1}{9}(1+\gamma_{i}) +\frac{1}{9}
\sum_{\alpha} \ln\[(t^{\alpha}+ \overline{t}^{\alpha})
|\eta(t^{\alpha})|^4\] \(\sum_{a} \gamma_{i}^{a} p_{ia}^{\alpha}
-2\sum_{jk} \gamma_{i}^{jk} p_{jk}^{\alpha} \) \right. \nonumber \\
 & & \left.  -\frac{1}{9} \sum_{a}
\gamma_{i}^{a} \ln(\tmu^2_{ia}) +\frac{2}{9} \sum_{jk} \gamma_{i}^{jk}
\ln(\tmu^2_{jk}) +\frac{\sin\theta}{k_{s\overline{s}}^{1/2}}\[
   \frac{1}{3\sqrt{3}} \( \sum_{a} \gamma_{i}^{a} g_{a}^{2}
   \cos\gamma_{S} - \frac{1}{2}
   \sum_{jk} \gamma_{i}^{jk} \(k_{s} e^{-i\gamma_{S}} +
   k_{\overline{s}} e^{i\gamma_{S}}\) \) \]
 \right. \nonumber \\
 & & \left. +\frac{\cos\theta}{3\sqrt{3}} \[ \sum_{\alpha}
    (t^{\alpha}+\overline{t}^{\alpha}) G^{\alpha}_{2}
   \Theta_{\alpha} \sum_{jk}
   \gamma_{i}^{jk} (n^{\alpha}_{i}+ n^{\alpha}_{j} + n^{\alpha}_{k}
   +1) \] \cos\gamma_{\alpha}  +\cos^{2}\theta \[ \frac{1}{3}\sum_{\alpha}n^{\alpha}_{i}
\Theta_{\alpha}^{2} \right.\right. \nonumber \\
 & & \left. \left. -\frac{1}{3} \sum_{\alpha} \Theta^{2}_{\alpha}
 \( \sum_{a} \gamma_{i}^{a} p_{ia}^{\alpha} +\sum_{jk} \gamma_{i}^{jk}
 p_{jk}^{\alpha} \) -\frac{1}{3} \sum_{a}\gamma_{i}^{a} \sum_{\alpha}
 n_{i}^{\alpha} \Theta^{2}_{\alpha} \ln(\tmu^2_{ia}) +\frac{1}{3} \sum_{jk}
 \gamma_{i}^{jk} \sum_{\alpha} (n_{j}^{\alpha}+n_{k}^{\alpha})
 \Theta^{2}_{\alpha} \ln(\tmu^2_{jk}) \right. \right. \nonumber \\
 & & \left. \left. -\sum_{\alpha} \ln\[(t^{\alpha}+ \overline{t}^{\alpha})
|\eta(t^{\alpha})|^4\] \(  -\frac{1}{3} \sum_{a} \gamma_{i}^{a}
p_{ia}^{\alpha} \sum_{\beta} n_{i}^{\beta} \Theta^{2}_{\beta} +
\frac{1}{3} \sum_{jk} \gamma_{i}^{jk}
p_{jk}^{\alpha} \sum_{\beta} (n_{j}^{\beta} +n_{k}^{\beta})
\Theta^{2}_{\beta} \right. \right. \right. \nonumber \\
& & \left. \left. \left. +\frac{1}{3} \sum_{jk} \gamma_{i}^{jk}
p_{jk}^{\alpha} \sum_{\beta}\sum_{\gamma}
   (t^{\beta}+\overline{t}^{\beta}) (t^{\gamma}+\overline{t}^{\gamma})
    G^{\beta}_{2} G^{\gamma}_{2}
   \Theta_{\beta}\Theta_{\gamma} (n^{\beta}_{i}+ n^{\beta}_{j} + n^{\beta}_{k}
   +1)(n^{\gamma}_{i}+ n^{\gamma}_{j} + n^{\gamma}_{k} +1) \)
   e^{-i(\gamma_{\beta}-\gamma_{\gamma})} 
 \right. \right. \nonumber \\
& & \left. \left. +\frac{1}{3} \sum_{jk} \gamma_{i}^{jk}
p_{jk}^{\alpha} \sum_{\alpha}\sum_{\beta} 
   (t^{\alpha}+\overline{t}^{\alpha}) (t^{\beta}+\overline{t}^{\beta})
   G^{\alpha}_{2} G^{\beta}_{2}
   \Theta_{\alpha}\Theta_{\beta} (n^{\beta}_{i}+ n^{\beta}_{j} + n^{\beta}_{k}
   +1) \cos(\gamma_{\beta}-\gamma_{\alpha}) \right. \right. \nonumber \\
& & \left. \left. + \frac{1}{3} \gamma_{i}^{jk}
    \sum_{\alpha}\sum_{\beta}
   (t^{\alpha}+\overline{t}^{\alpha}) (t^{\beta}+\overline{t}^{\beta})
    G^{\alpha}_{2} G^{\beta}_{2}
   \Theta_{\alpha}\Theta_{\beta} (n^{\alpha}_{i}+ n^{\alpha}_{j} + n^{\alpha}_{k}
   +1) (n^{\beta}_{i}+ n^{\beta}_{j} + n^{\beta}_{k} +1) \ln(\tmu^2_{jk})
   e^{-i(\gamma_{\alpha}-\gamma_{\beta})} 
   \] \right.  \nonumber \\ 
& & \left. +\frac{\sin\theta\cos\theta}{k_{s\overline{s}}^{1/2}} \[
   -\frac{1}{6}\sum_{\alpha} 
   (t^{\alpha}+\overline{t}^{\alpha}) G^{\alpha}_{2} \Theta_{\alpha} \sum_{jk}
   \gamma_{i}^{jk} p^{\alpha}_{jk} \(k_{s} e^{-i(\gamma_{S} -
     \gamma_{\alpha})} +k_{\overline{s}} e^{i(\gamma_{S} - \gamma_{\alpha})}\)
     \right. \right. \nonumber
\\
& & \left. \left. +\frac{1}{3}\sum_{\alpha} 
   (t^{\alpha}+\overline{t}^{\alpha}) G^{\alpha}_{2} \Theta_{\alpha} \sum_{a}
   g_{a}^{2} \gamma_{i}^{a} p_{ia}^{\alpha} \cos(\gamma_{S} -
   \gamma_{\alpha})   \right. \right. \nonumber \\
& & \left. \left.  +\frac{1}{3} \sum_{\alpha} \ln\[ (t^{\alpha}+
    \overline{t}^{\alpha}) |\eta(t^{\alpha})|^4\] \sum_{jk}
    \gamma_{i}^{jk} p_{jk}^{\alpha} \(k_{s} e^{-i(\gamma_{\beta} -
      \gamma_{S})}  + k_{\overline{s}} e^{i(\gamma_{\beta} - \gamma_{S})}\)
    \sum_{\beta}
   (t^{\beta}+\overline{t}^{\beta})  G^{\beta}_{2} \Theta_{\beta} (n^{\beta}_{i}+
   n^{\beta}_{j} + n^{\beta}_{k} +1) \right. \right. \nonumber \\
 & & \left. \left. + \frac{1}{6} \sum_{jk} \gamma_{i}^{jk} \sum_{\alpha}
      (t^{\alpha}+\overline{t}^{\alpha}) G^{\alpha}_{2} \Theta_{\alpha}
     (n^{\alpha}_{i}+ n^{\alpha}_{j} + n^{\alpha}_{k} +1)\(
     \partial_{\overline{s}} \ln(\tmu^2_{jk}) + k_{\overline{s}}
     \ln(\tmu^2_{jk}) \) e^{i(\gamma_{S} - \gamma_{\alpha})}  \right.
   \right. \nonumber \\ 
 & & \left. \left. + \frac{1}{6} \sum_{jk} \gamma_{i}^{jk} \sum_{\alpha}
      (t^{\alpha}+\overline{t}^{\alpha}) G^{\alpha}_{2} \Theta_{\alpha}
     (n^{\alpha}_{i}+ n^{\alpha}_{j} + n^{\alpha}_{k} +1)\(
     \partial_{s} \ln(\tmu^2_{jk}) + k_{s} \ln(\tmu^2_{jk}) \)
     e^{-i(\gamma_{S} - \gamma_{\alpha})} \]
   \right. \nonumber \\ 
& & \left. + \frac{\sin^{2}\theta}{k_{s\overline{s}}} \[ -\frac{1}{4}
    \sum_{a} g_{a}^{4} \gamma_{i}^{a} \ln(\tmu^2_{ia}) - \frac{1}{3} \sum_{a}
     \gamma_{i}^{a} \partial_{s} \partial_{\overline{s}} \ln(\tmu^2_{ia})
    +\frac{1}{3}\sum_{a} \gamma_{i}^{a} g_{a}^{2} \(
    \partial_{s} \ln(\tmu^2_{ia}) + \partial_{\overline{s}} \ln(\tmu^2_{ia}) \)
  \right. \right. \nonumber \\ 
& & \left. \left. -\sum_{jk} \gamma_{i}^{jk}
    \( \frac{1}{3}(k_{s}k_{\overline{s}} +2k_{s\overline{s}})
    \ln(\tmu^2_{jk}) +\frac{1}{3}\partial_{s} \partial_{\overline{s}}
    \ln(\tmu^2_{jk}) +\frac{1}{2}(k_{s}\partial_{\overline{s}}\ln(\tmu^2_{jk}) +
    k_{\overline{s}}\partial_{s}\ln(\tmu^2_{jk})) \)
  \right. \right. \nonumber \\ 
& & \left. \left. -\sum_{\alpha} \ln\[(t^{\alpha}+ \overline{t}^{\alpha})
|\eta(t^{\alpha})|^4\] \( \frac{1}{4} \sum_{a} g_{a}^{4}
\gamma_{i}^{a} p_{ia}^{\alpha} +\frac{1}{3} \sum_{jk} \gamma_{i}^{jk}
p_{jk}^{\alpha} k_{s} k_{\overline{s}} \) \] \rbr.
\label{massfull}
\end{eqnarray}
\end{footnotesize}

\end{document}